\pdfoutput=1
\documentclass[journal]{IEEEtran}
\usepackage {graphicx,alltt,epsf}
\usepackage{times, amsmath, amssymb, epsfig}
\usepackage{multirow}
\usepackage{hyperref}

\begin{document}
\title{Maximizing the Signal-to-Alias Ratio in Non-Uniform Filter Banks for Acoustic Echo Cancellation} 
\author{R.~C.~Nongpiur,~\IEEEmembership{Member,~IEEE,}
        and~D.~J.~Shpak,~\IEEEmembership{Senior Member,~IEEE} 
\thanks{R.~C.~Nongpiur and D.~J.~Shpak are with the Department
of Electrical and Computer Engineering, University of Victoria, Victoria,
BC, Canada V8W 3P6 e-mail: rnongpiu@ece.uvic.ca; dshpak@ece.uvic.ca}
\thanks{Manuscript submitted May 2011.}}
\maketitle
\begin{abstract}
A new method for designing non-uniform filter-banks for acoustic echo cancellation is proposed. In the method, the analysis prototype filter design is framed as a convex optimization problem that maximizes the signal-to-alias ratio (SAR) in the analysis banks. Since each sub-band has a different bandwidth, the contribution to the overall SAR from each analysis bank is taken into account during optimization. To increase the degrees of freedom during optimization, no constraints are imposed on the phase or group delay of the filters; at the same time, low delay is achieved by ensuring that the resulting filters are minimum phase. Experimental results show that the filter bank designed using the proposed method results in a sub-band adaptive filter with a much better echo return loss enhancement (ERLE) when compared with existing design methods.
\end{abstract}

\begin{IEEEkeywords}
acoustic echo cancellation, non-uniform filter-banks, sub-band adaptive filter
\end{IEEEkeywords}

\section{Introduction}
Non-uniform filter banks are of interest in speech processing applications since they can be used to exploit the perceptual properties of the human ear~\cite{zwicker}. A well known and efficient technique to realize a non-uniform filter-bank is the all-pass transformed polyphase filter-bank~\cite{oppenheim}-\cite{doblinger}, where the delay elements of the input and output delay chains are replaced by first-order all-pass filters, as shown in Fig. 1. Such a warped filter bank has been found to be beneficial in applications such as speech enhancement and beamforming~\cite{gulzow, nordholm}. In addition, the warped filter banks also involve much lower delay and complexity in comparison to non-uniform filter banks realized by a tree structure~\cite{gulzow}. Since most hands-free and speech enhancement systems are coupled with an acoustic echo canceller~\cite{gerhard}, it is important that the analysis and synthesis filter banks are optimized for echo cancellation. 

Other realizations of non-uniform filter structures are obtained by joining two or more uniform filter bank structures of different bandwidths by transitions banks~\cite{princen}-\cite{cvetkovic}, or by combining a subset of varying numbers of subbands of a uniform filter bank~\cite{mccloud1}-\cite{lee}. In~\cite{batalheiro, petraglia2}  critically sampled non-uniform filter banks for adaptive filtering are realized by incorporating extra filters in between the non-uniform sub-bands to cancel the aliasing.

In echo cancellation for speech signals, cancellation of low-frequency echoes is most critical for two important reasons~\cite{gerhard}. The first is because most of the speech energy is distributed in the low-frequency end of the audio spectrum. The second is due to room acoustics: in a typical room environment the higher-frequency components of an audio signal are more easily absorbed by the materials in the room (walls, carpets, curtains, etc.) and, as a result, the lower frequency sub-bands require much longer adaptive filter lengths to cancel the echoes. Consequently, by using non-uniform filter banks that have bandwidths that increase with frequency, the convergence rate of the lower sub-bands can be improved significantly thereby resulting in more effective cancellation of the low-frequency echoes.

The use of sub-band adaptive filters in acoustic echo cancellation has been quite popular, especially when the impulse response is very long, due to their fast convergence rate and low computational complexity in comparison to full-band adaptive filters~\cite{kellerman}-\cite{wilbur}. In sub-band echo cancellation, one of the critical aspects of filter bank design is the minimization of the aliasing component during the analysis stage, as aliasing disturbs the convergence process of the adaptive filter. It is well known that aliasing in the sub-band signals caused by finite stop-band attenuation influences the MMSE~\cite{kellerman}, \cite{weiss}-\cite{slock}. Efforts to quantify the MMSE via aliasing have been carried out in~\cite{weiss}, \cite{petraglia}-\cite{morgan}.

In~\cite{kliewer}-\cite{vary3} non-uniform filter-banks were designed with emphasis on near-perfect reconstruction (NPR) of the analysis-synthesis system. Although these designs are useful in applications such as speech coding, they usually do not work well in adaptive filtering since the signal components in the adjacent bands that are required for NPR are often severely modified by an adaptive filter. In~\cite{nordholm, vo} non-uniform filter-banks that minimize aliasing during the analysis stage were developed for beamforming and speech processing applications. With this approach, a linear phase constraint is imposed on both the analysis and synthesis prototype filters, and the filter group-delay, which may not be optimal, must be specified. 

In~\cite{rnongpiur1}, we framed the design method without phase constraints on the filters, which increases the degrees of freedom during optimization, and, in turn, improves the aliasing-suppression performance of the filters. Then, in~\cite{rnongpiur2} we modified the objective function so that overall signal-to-alias ratio (SAR) is maximized. 
\begin{figure}[tbp]
\begin{center}
\includegraphics[width=0.45\textwidth]{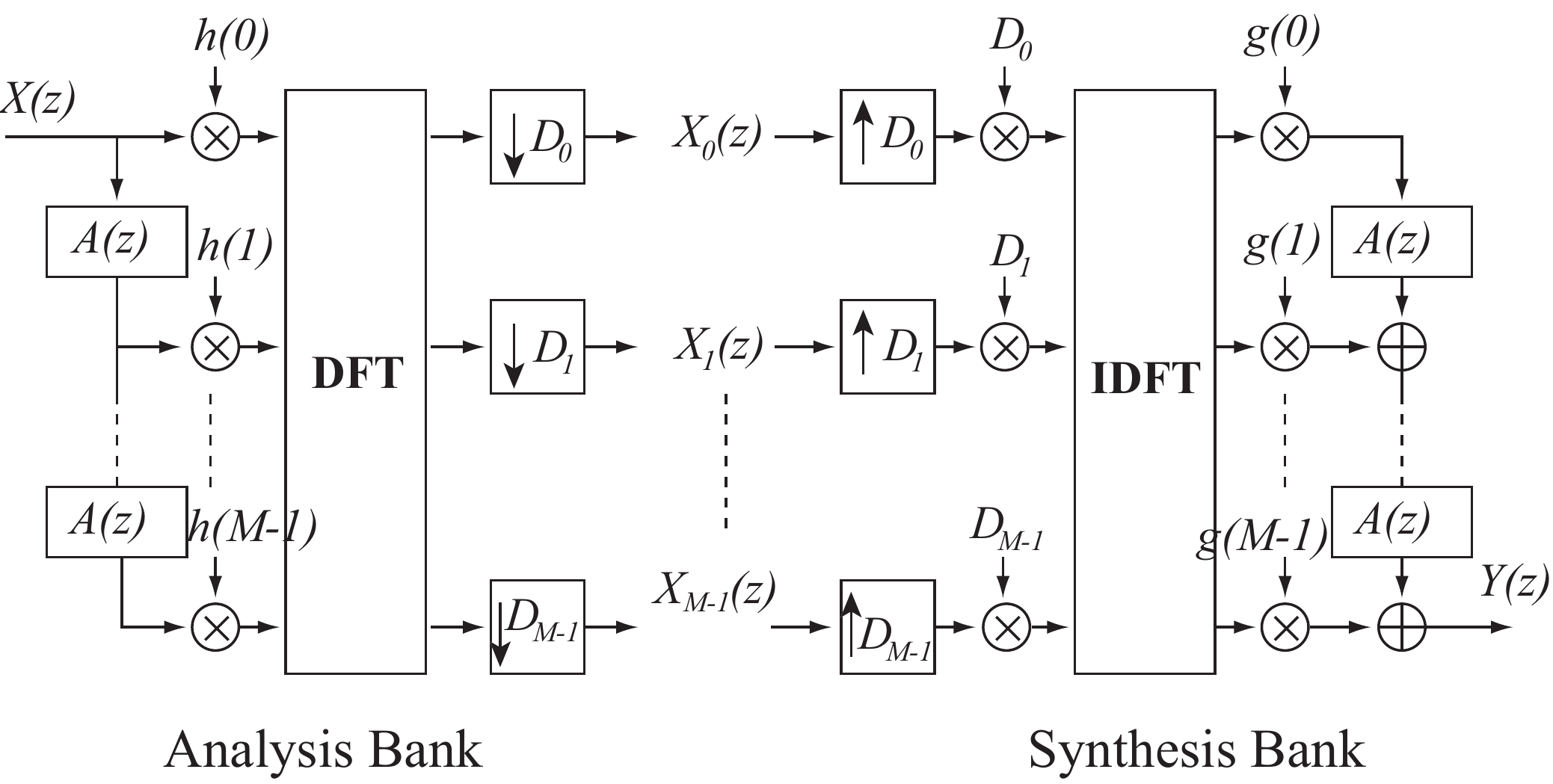}
\caption{The DFT-transformed non-uniform polyphase filter-bank with variable decimation factors.}
\label{fig:f1}
\end{center}
\end{figure}
The SAR characterizes the factor by which the error signal power can be reduced by adaptive filtering and is equivalent to the widely used ERLE quality measure~\cite{weiss}. Since each subband in a non-uniform filter bank has a different bandwidth, the contribution to SAR from each subband will be different. Consequently, to ensure that the overall SAR is maximized the contribution from each of the analysis banks, as well as the PSDs of the input signal, $x(n)$, and the unknown system, $s(n)$, are taken into account during optimization in~\cite{rnongpiur2}. 
In this paper, we extend and improve on the method developed in~\cite{rnongpiur2}. We describe how the maximization of SAR across the subbands leads to an increase in ERLE performance; then, we formulate a convex optimization problem so that the SAR is maximized across the subbands. Experimental results show that the filter bank designed using the proposed method results in a much lower ERLE when compared to existing design methods.

 The paper is organized as follows. Section II describes the non-uniform filter-bank implementation while Section III describes the subband adaptive filter. In Sections IV and V, the design of the analysis and synthesis prototype filters, respectively, are discussed. In Section VI, experimental results are presented to show the effectiveness of the proposed approach. Conclusions are drawn in Section VII.
\section{The Non-Uniform DFT Filter Bank}
The non-uniform filter bank in Fig.~1 is a generalization of the uniform DFT filter bank where the delay element, $z^{-1}$, is replaced by a first-order allpass filter, $A(z)$, of the form
\begin{equation}
A(z) = \frac{\mu z + 1}{z+ \mu} \mbox{ where } |\mu| < 1
\label{eq:1}
\end{equation}
Using an $M$-point DFT analysis bank, the transfer function, $H_i(z)$, and z-domain output signal, $X_i(z)$,  of the $i$th analysis subband filter are given by
\begin{eqnarray}
H_{i}(z) = \sum_{n=0}^{M-1} h(n) W_M^{ni} A(z)^n \label{eq:2} \\
X_{i}(z) = \frac{1}{D_i} \sum_{d=0}^{D_i-1} X(z^{\frac{1}{D_i}}W_{D_i}^d) H_i(z^{\frac{1}{D_i}} W_{D_i}^d) \label{eq:3}
\end{eqnarray}
where $h(n)$ is the analysis prototype filter, $X(z)$ is the z-domain input signal, $D_i$ is the downsampling factor in the $i$th sub-band, and $W_M = e^{-j2\pi/M}$ is the complex modulating factor. The corresponding synthesis bank is an $M$-point inverse-DFT, with the $i$th synthesis subband filter given by
 \begin{equation}
G_i(z) = \sum_{n=0}^{M-1} g(n) W_M^{-ni} A(z)^{M-n-1}
\label{eq:4}
\end{equation}
where $g(n)$ is the synthesis prototype filter. The overall input-output relationship for the analysis-synthesis system can be expressed as
 \begin{equation}
Y(z) = \sum_{i=0}^{M-1} G_i(z) \sum_{d=0}^{D_i-1} X(z W^d_{D_i}) H_i(zW_{D_i}^d)
\label{eq:5}
\end{equation}
In general, the input-output transfer function of the analysis-synthesis system is a linear, periodically time varying system with period equal to the maximum downsampling factor $D_{max}$. Therefore, to account for this behaviour, the overall transfer function is computed by using a sequence of $D_{max}$ time-shifted impulses as input and given by 
 \begin{equation}
T_l(z) = \frac{Y(z)}{z^{-l}} = \sum_{i=0}^{M-1} G_i(z)  \sum_{d=0}^{D_i-1} W_{D_i}^{-dl} H_i(zW_{D_i}^d)
\label{eq:6}
\end{equation}
where we have assumed $X(z) = z^{-l}$ for $l \ \ \epsilon$  $\{0, 1, \ldots,$ $ (D_{max}-1)\}$, and 
 \begin{equation}
D_{max} = \max_i D_i
\end{equation}
As such, by replacing the delay element $z^{-1}$ by the all-pass filter $A(z)$, the frequency response of the filter at frequency $\omega$ is mapped into frequency $\Omega$, given by~\cite{oppenheim}
\begin{equation}
\Omega = \phi(\omega) = \tan^{-1} \left[\frac{(1-\mu^2)\sin \omega}{(1+\mu^2)\cos\omega+2\mu}   \right]
\end{equation}
Consequently, the $i$th subband filter $H_i(z)$ will lie between frequencies $\Omega_l^{(i)}$ and $\Omega_h^{(i)}$ where
\begin{eqnarray}
\Omega_l^{(i)} & = & D_i\phi(\omega_c^{(i)}  - x) \label{omega_l}\\
\Omega_h^{(i)} & = & D_i\phi(\omega_c^{(i)}  + x) \label{omega_h}\\
\omega_c^{(i)} & = & \frac{2\pi i}{M} \\
\mbox{and  } \Omega_h^{(i)} & = & \Omega_l^{(i)} + 2\pi
\end{eqnarray} 
Parameters $\Omega_l^{(i)}$ and $\Omega_h^{(i)}$ can be obtained by solving for $x$ using a simple line search optimization algorithm on the convex function
\begin{equation}
\mbox{minimize } \left(\phi(\omega_c^{(i)}  + x)-\phi(\omega_c^{(i)}  - x) - 2\pi/D_i \right)^2
\label{lineSearch}
\end{equation}
where $x$ is the optimization variable. In Section III, frequencies $\Omega_l^{(i)}$ and $\Omega_h^{(i)}$ will be used as integration limits when computing the aliasing power in the subband filter $H_i(z)$.
\section{The Sub-band Adaptive Filter}
\begin{figure}[tbp]
\begin{center}
\includegraphics[width=0.45\textwidth]{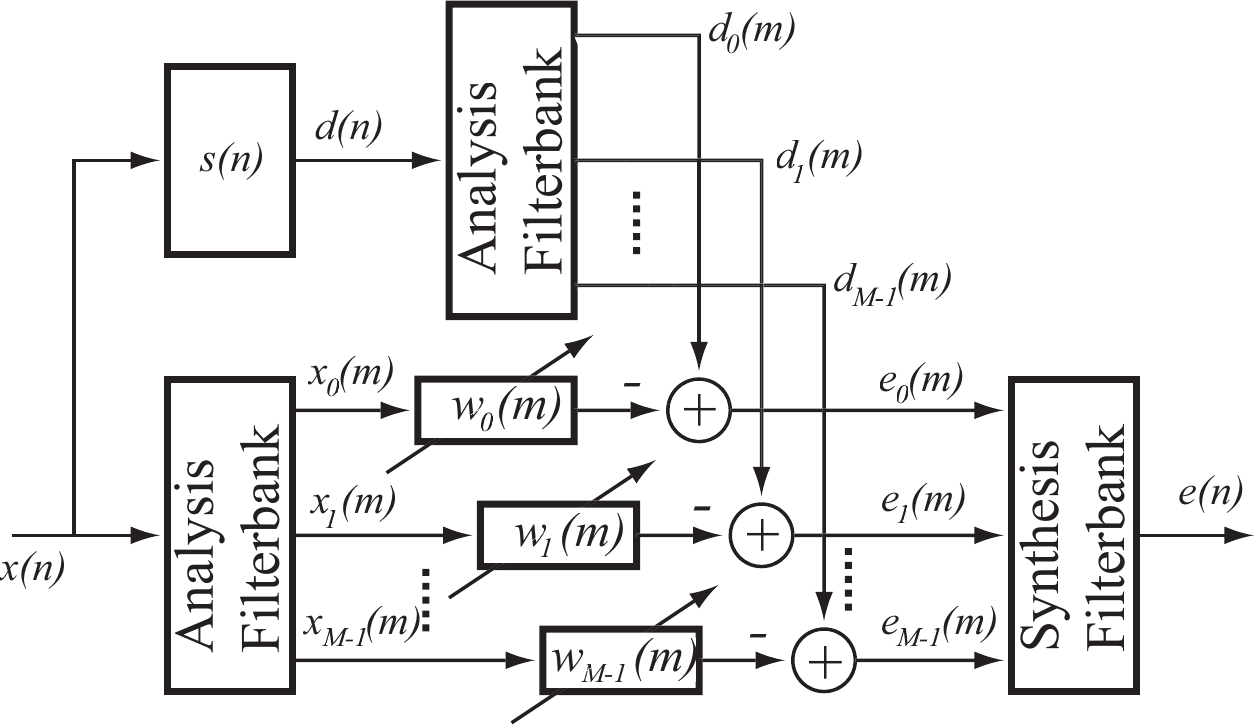}
\caption{The subband adaptive filter}
\label{adaptiveFilter}
\end{center}
\end{figure}
In Fig.~2 the sub-band adaptive filter structure is shown. As can be seen, the input signal $x(n)$ and desired signal $d(n)$ are split into $M$ subbands by analysis filter banks. The resulting subband signals $x_i(m)$ and $d_i(m)$ in the $i$th subband are adapted independently of the other subband signals. The resulting errors $e_i(m)$ from each of the subbands are then recombined to form the fullband error signal $e(n)$. 

In most adaptive filtering applications, the signal $x(n)$ is represented as a stochastic signal with known power spectral density (PSD) $P_{xx}$. To characterize $x(n)$ by a spectrum rather than a PSD we represent $x(n)$ as the output of a source model $F(e^{j\omega})$ which is excited by a white noise signal $u(n)$ of unit variance~\cite{weiss}. Using the spectrum representation, the PSD of $x(n)$ is given by 
\begin{equation}
P_{xx}(e^{j\omega}) = |F(e^{j\omega}|^2
\end{equation}
From Fig.~2, the $i$th desired signal will be a combination of the unknown system $S(e^{j\omega})$, the subband filter $H_i(e^{j\omega})$, and source model for the input signal $F(e^{j\omega})$ given by
\begin{equation}
L_i(e^{j\omega}) = F(e^{j\omega}) S(e^{j\omega}) H_i(e^{j\omega})
\label{model}
\end{equation}

\subsection{Relationship between ERLE and SAR}
For a fullband adaptive filter with filter coefficients $\hat{s}_{(i)}(n)$, the ERLE of an echo canceller is defined as
\begin{equation}
\mbox{ERLE}(n) = \frac{E[d^2(n)]}{E[((d(n)-\hat{d}(n))^2}
\end{equation}
where $\hat{d}(n)$ is the adaptive filter estimate of the desired signal $d(n)$ and is given by
\begin{equation}
\hat{d}(n) = \sum_{i=0}^{N-1} \hat{s}_{(i)}(n)x(n-i)
\end{equation}
If we assume, for simplicity, a stationary white noise input signal $x(n)$, the ERLE can be expressed as
\begin{equation}
\begin{split}
& \mbox{ERLE}(n) =  \\
& \frac{E[x^2(n)] \displaystyle \sum_{i=0}^\infty s_{(i)}^2(n)}{E[x^2(n)]\left( \displaystyle \sum_{i=0}^\infty s_{(i)}^2(n) - 2 \displaystyle \sum_{i=0}^{N-1}s_{(i)}(n)\hat{s}_{(i)}(n) + \displaystyle \sum_{i=0}^{N-1} \hat{s}_{(i)}^2(n) \right)} 
\end{split}
\end{equation}
Assuming perfect match of the $N$ coefficients of the adaptive filter, so that 
\begin{equation}
\hat{s}_{(i)}(n) = s_{(i)}(n) \mbox{  for $0 \leq i < N$}
\end{equation}
the upper bound of the ERLE simplifies to
\begin{equation}
\mbox{ERLE}_{max}(n, N) = \frac{\sum_{i=0}^\infty s_{(i)}^2(n)}{\sum_{i=N}^\infty s_{(i)}^2(n)}
\label{erleSimpl}
\end{equation}
As can be seen from (\ref{erleSimpl}), if the filter length is made long enough the ERLE can be made arbitrarily small in a full-band adaptive filter. 

In a subband adaptive filter, however, the ERLE is dependent not only on the length of the adaptive filter, but also on the amount of aliasing power present after analysis and synthesis. If the length of each subband adaptive filter is made sufficiently long, the ERLE will then be dependent only on the power ratio between the desired signal and the steady state error due to aliasing, or SAR~\cite{weiss}. Therefore, for a sub-band adaptive filter with sufficiently long sub-band adaptive filters so that the impulse response of the unknown system is adequately modelled, we have
\begin{equation}
\mbox{ERLE} \propto \mbox{SAR}
\end{equation}
To compute the SAR, we use the approximation in~\cite{weiss} and extend it to the non-uniform filter bank case, giving
\begin{equation}
\mbox{SAR} \approx \frac{\displaystyle \sum_{i=0}^{M-1}\sigma^2_i}{\displaystyle \sum_{i=0}^{M-1}(\sigma_i^{(a)})^2}
\label{overallSAR}
\end{equation} 
where
\begin{equation}
\sigma_i^2 = \frac{1}{2\pi} \int_0^{2\pi} \left|\sum_{d=0}^{D_i-1} L_i \left( e^{j(\omega - 2\pi d)/D_i} \right) \right|^2 d\omega \ \ \ \ \mbox{and}
\label{sig1}
\end{equation}
\begin{equation}
(\sigma_i^{(a)})^2 = \frac{1}{2\pi} \int_{\Omega_l^{(i)}}^{\Omega_h^{(i)}} \left|\sum_{d=1}^{D_i-1} L_i \left( e^{j(\omega - 2\pi d)/D_i} \right) \right|^2 d\omega
\label{sig2}
\end{equation}
The SAR in each sub-band is given by
\begin{equation}
\mbox{SAR}_i = \frac{\sigma^2_i}{(\sigma_i^{(a)})^2} 
\label{subbandSAR}
\end{equation}
Equations (\ref{sig1}) and (\ref{sig2}) can be simplified if we exchange summation and squaring by ignoring the mixed product terms in the source model, which is justified if the unknown system is comprised of statistically independent frequency components ~\cite{weiss}. Therefore, $\sigma_i^2$ and $(\sigma_i^{(a)})^2$ become
\begin{eqnarray}
\sigma_i^2 \ & \approx & \frac{1}{2\pi} \int_0^{2\pi} \sum_{d=0}^{D_i-1} \left| L_i \left( e^{j(\omega - 2\pi d)/D_i} \right) \right|^2 d\omega \notag \\
& = & \frac{D_i}{2\pi} \int_0^{2\pi} \left| L_i(e^{j\omega}) \right|^2 d\omega \ \ \ \ \mbox{and}
\label{simpSig1}
\end{eqnarray}
\begin{equation}
(\sigma_i^{(a)})^2 = \frac{1}{2\pi} \int_{\Omega_l^{(i)}}^{\Omega_h^{(i)}} \sum_{d=1}^{D_i-1} \left| L_i \left( e^{j(\omega - 2\pi d)/D_i} \right) \right|^2 d\omega
\label{simpSig2}
\end{equation}
Using (\ref{model}), $|L_i(e^{j\omega})|^2$ above can be expanded as
\begin{equation}
|L_i(e^{j\omega})|^2 =  P_{xx}(e^{j\omega}) |S(e^{j\omega})|^2 |H_i(e^{j\omega})|^2
\label{modelExpand}
\end{equation}
If $P_{xx}(e^{j\omega})$ and the average power spectrum of $S(e^{j\omega})$ are not readily available, we can simplify further by setting
\begin{equation}
P_{xx}(e^{j\omega}) = |S(e^{j\omega})|^2 = 1
\label{whiteNoiseAssump}
\end{equation}
\section{Analysis Filter Bank Design}
To design the analysis filter with no phase constraint, the square of the magnitude of the frequency response is used. To this end, from (\ref{eq:2}) we get
\begin{equation}
\begin{split}
|H_i(e^{j\omega})|^2 & = \sum_{k=0}^{M-1} h(k)W_M^{ki} A(e^{j\omega})^k \sum_{k=0}^{M-1} h(k)W_M^{-ki}A(e^{j\omega})^{-k} \\
& = \sum_{k=-(M-1)}^{M-1}c(k) W_M^{ki}A(e^{j\omega})^k \ \ \ \ \mbox{where}
\end{split}
\label{eq:7}
\end{equation}
\begin{equation}
c(k) = c(-k) \mbox{ for } k = 0, 1, \ldots, M-1
\label{magCoeff}
\end{equation}
The magnitude-squared function in (\ref{eq:7}) can be further simplified as
\begin{equation}
|H_i(e^{j\omega})|^2 = c(0) + 2 \sum_{k=1}^{M-1} c(k) \Re [W_M^{ik} A(e^{j\omega})^{k}]
\label{eq:8}
\end{equation}
where $\Re[\mathbf{\cdot}] $ gives the real part of a complex number. To get the minimum phase prototype filter $h_{mp}(k)$ given $c(k)$, we use the property that any two filters having identical magnitude response when $A(z)=z^{-1}$ will have identical magnitude response for any $A(z)$; as such, we first compute the real cepstrum, $\kappa(n)$, of $\hat{H}_0(e^{j\omega}) = H_0(e^{j\omega})$ $|_{A(z)=z^{-1}}$ using the expression
\begin{equation}
\kappa(n) = \frac{1}{\pi}\int_{-\pi}^{\pi} \log{|\hat{H}_0(e^{j\omega})|} e^{j\omega n} d\omega
\label{eq:9}
\end{equation}
and then compute $h_{mp}(k)$ from $\kappa[n]$ by taking the inverse cepstrum~\cite{algobook}.
\subsection{The Optimization Problem}
The prototype filter is designed by minimizing the SAR across all of the analysis subbands. To this end, we solve the optimization problem:
\begin{eqnarray}
\mbox{minimize } & &  \sum_{i=0}^{M} (\sigma_i^{(a)})^2  \label{genOptm} \\
\mbox{subject to: } & & \sum_{i=0}^{M} \sigma_i^2 = \mbox{constant} \notag 
\end{eqnarray}
with the prototype filter-magnitude coefficients as the optimization variables. To obtain the global minimum, we frame the optimization as a convex optimization problem, which is done by ensuring that the cost function is convex and the equality constraint is affine~\cite{wslu}. 

By using the coefficients of the magnitude squared coefficients in (\ref{magCoeff}) as the optimization variable and combining (\ref{simpSig2}), (\ref{modelExpand}), and (\ref{eq:8}) we can express the cost function in affine form, which is convex, as
\begin{equation}
\begin{split}
\sum_{i=0}^{M-1} (\sigma_i^{(a)})^2  & = \frac{1}{2\pi} \sum_{i=0}^{M-1}  \int_{\Omega_l^{(i)}}^{\Omega_h^{(i)}} \sum_{d=1}^{D_i-1} \left| L_i \left( e^{j(\omega - 2\pi d)/D_i} \right) \right|^2 d\omega  \\
& \approx  \frac{1}{N}\sum_{i=0}^{M-1} \sum_{p = 0}^{N-1} \sum_{d=1}^{D_i-1} \left| L_i \left( e^{j(\omega_p - 2\pi d)/D_i} \right) \right|^2 \\
& =  \mathbf{1}^T\mathbf{A}\mathbf{c} 
\label{anaCostFun}
\end{split}
\end{equation}
where $\omega_p$ $\in$ $[\Omega_l^{(i)}, \Omega_h^{(i)}]$, $\mathbf{1}$ $\in$ $\mathbf{R}^{MN}$,
\begin{eqnarray}
\mathbf{A} &=& \left[ [\mathbf{A}^{(0)}]^T, \cdots, [\mathbf{A}^{(M-1)}]^T \right]^T \notag \\
\mathbf{A}^{(i)} &=& \frac{2}{N}\left[ \begin{matrix}
a_{00}^{(i)} & \cdots & a_{0(M-1)}^{(i)}  \cr
\vdots & \vdots & \vdots \cr
a_{(N-1)0}^{(i)}  & \cdots & a_{(N-1)(M-1)}^{(i)}  \cr
\end{matrix}
\right ],  \notag \\
a_{pq}^{(i)} \hspace{-0.1in} &=&  \hspace{-0.1in}\begin{cases}
\frac{1}{2}P_{xx}(e^{\frac{\omega_p}{D_i}}W^d_{D_i}) |S(e^{\frac{\omega_p}{D_i}}W^d_{D_i})|^2 (D_i-1) & \text{if $q = 0$}, \notag \\ \\
P_{xx}(e^{\frac{\omega_p}{D_i}}W^d_{D_i}) |S(e^{\frac{\omega_p}{D_i}}W^d_{D_i})|^2 \times & \notag \\ 
\ \ \ \ \ \ \ \ \ \ \ \ \displaystyle \sum_{d=1}^{D_i-1} \Re[W_M^{iq}A(e^{\frac{\omega_p}{D_i}}W^d_{D_i})^q] & \text{else}.
\end{cases} \notag \\
\mathbf{c} &=& [c_0, c_1, \ldots, c_{M-1}]^T, \notag
\label{eq:11}
\end{eqnarray}
In a similar manner, the left hand side of the equality constraint in (\ref{genOptm}) can be expressed in affine form as 
\begin{equation}
\begin{split}
\sum_{i=0}^{M-1} \sigma_i^2 & = \frac{1}{2\pi} \sum_{i=0}^{M-1} D_i \int_0^{2\pi} \left| L_i(e^{j\omega}) \right|^2 d\omega   \\
& \approx \frac{1}{N}  \sum_{i=0}^{M-1} D_i \sum_{p = 0}^{N-1} \left| L_i(e^{j\omega_p}) \right|^2, \ \ \ \ \ \omega_p \ \epsilon \ \ [-\pi, \pi] \\
& = \mathbf{1}^T\mathbf{B}\mathbf{c} 
\end{split}
\label{eq:eqConstr}
\end{equation}
where
\begin{eqnarray}
\mathbf{B} &=&  \left[ [\mathbf{B}^{(0)}]^T, \cdots, [\mathbf{B}^{(M-1)}]^T \right]^T \notag \\  
\mathbf{B}^{(i)} &=& \frac{2D_i}{N}\left[ \begin{matrix}
b_{00}^{(i)} & \cdots & b_{0(M-1)}^{(i)}  \cr
\vdots & \vdots & \vdots \cr
b_{(N-1)0}^{(i)}  & \cdots & b_{(N-1)(M-1)}^{(i)}  \cr
\end{matrix}
\right ], \mbox{ and} \notag \\
b_{pq}^{(i)} & = & \begin{cases}
\frac{1}{2}P_{xx}(e^{j\omega_p}) |S(e^{j\omega_p})|^2 & \text{if $q = 0$}, \notag \\
P_{xx}(e^{j\omega_p}) |S(e^{j\omega_p})|^2 \Re[W_M^{ki} A(e^{j\omega_p})^q ] & \text{else}. \notag
\end{cases}
\label{eq:13}
\end{eqnarray}
Thus, we solve the following linear optimization problem:
\begin{eqnarray}
\mbox{minimize } & &  \mathbf{1}^T\mathbf{A}\mathbf{c} \label{genOptmLinear} \\
\mbox{subject to: } & & \mathbf{1}^T\mathbf{B}\mathbf{c} = \mbox{constant}  \notag \\
& & \mathbf{B}\mathbf{c} > \mathbf{0}
\end{eqnarray}
where $\mathbf{0}$ $\in$ $\mathbf{R}^{MN}$. The inequality constraint is a positivity constraint to ensure that the magnitude always remains positive. 

Once we obtain the optimal magnitude filter coefficients, $\mathbf{c}_{opt}$, we compute its cepstrum using (\ref{eq:9}) and then recover the minimum-phase filter coefficients, $h_{mp}(k)$, of the prototype filter.
\section{Synthesis Filter Bank Design}
The transfer function $T_l(z)$ in (\ref{eq:6}) can be divided into two signal components, the desired signal component, $T_d$, and the aliased signal component $T_a$; that is,
\begin{equation}
T_l(e^{j\omega}) = T_d(e^{j\omega}) + T_a(e^{j\omega}, l),\ \ \ \ \mbox{where}
\label{eq:18}
\end{equation}
\begin{equation}
\begin{split}
T_d(e^{j\omega}) & = \sum_{i=0}^{M-1} H_i(e^{j\omega})G_i(e^{j\omega}) \\
& = \sum_{i=0}^{M-1} \sum_{n=0}^{M-1} h(n) W_M^{ni} A(e^{j\omega})^n \\
& \hspace{20 mm}\sum_{m=0}^{M-1} g(m) W_M^{-mi} A(e^{j\omega})^{M-m-1} \\
& =  A(e^{j\omega})^{M-1} \sum_{n=0}^{M-1} h(n)g(n),\ \ \ \ \mbox{and}
\end{split}
\label{eq:19}
\end{equation}
\begin{equation}
\begin{split}
T_a(e^{j\omega}, l) & = \sum_{i=0}^{M-1} G_i(e^{j\omega}) \sum_{d=1}^{D_i-1} W_{D_i}^{-dl}   H_i(e^{j\omega}W_{D_i}^d)
\end{split}
\label{eq:20}
\end{equation}
The cost function for the aliasing power is taken as the power sum of (\ref{eq:20}) summed across the spectrum for all combinations of $l$, given by
\begin{equation}
\begin{split}
\Upsilon & = \sum_{l=0}^{D_{max}-1} \sum_{n=0}^{N-1} \left|    \sum_{i=0}^{M-1} \sum_{d=1}^{D_i-1} W_{D_i}^{-dl}  G_i(e^{j\omega_n})  H_i(e^{j\omega_n}W_{D_i}^d) \right|^2 \\
& = \mathbf{g}^T \mathbf{S} \ \mathbf{g}\\
\end{split}
\label{eq:21}
\end{equation}
where
\begin{equation}
\begin{split}
\mathbf{g} & = [g(0), g(1), \ldots, g(M-1)]^T \\
\mathbf{S} & = \sum_{l=0}^{D_{max}-1} \sum_{n=0}^{N-1} \mathbf{Q}_{ln} \ \mathbf{Q}^H_{ln} \\
\mathbf{Q}_{ln} & = [q_{ln}(0), q_{ln}(1), \ldots, q_{ln}(M-1)]^T \\
q_{ln}(k) & = \sum_{i=0}^{M-1} \sum_{d=1}^{D_i-1} W_{D_i}^{-dl}  W_M^{-ki} H_i(e^{j\omega_n}W_{D_i}^d)A(e^{j\omega_n})^{M-k-1}
\end{split}
\label{eq:21a}
\end{equation}
The synthesis filter is designed by minimizing the aliasing cost function, $\Upsilon$, subject to the constraint that the magnitude of $T_d(e^{j\omega})$ is unity; as a consequence, we solve the quadratic optimization problem:
\begin{eqnarray}
\mbox{minimize } & & \mathbf{g}^T \mathbf{S} \ \mathbf{g} + \delta \mathbf{g}^T \mathbf{g} \label{eq:22} \\
\mbox{subject to: } & & \mathbf{h}^T\mathbf{g} = 1
\end{eqnarray}
where $\mathbf{g}$ $\epsilon$ $\mathbf{R}^M$ is the optimization variable, $\delta$ is a small positive number, and $\mathbf{h} = [h(0), h(1), \ldots,$ $ h(M-1)]^T$. The term $\delta \mathbf{g}^T \mathbf{g}$ in (\ref{eq:22}) is a regularization parameter that is introduced in case the matrix $\mathbf{S}$ is ill-conditioned; for example, this may happen when some of the coefficients in $h(n)$ are 0.

If we assume that the magnitude of the aliased signal component, $T_a(e^{j\omega}, l)$, is adequately minimized, the frequency response of the analysis-synthesis system is dependent only on $T_d(e^{j\omega})$. Therefore, from (\ref{eq:19}), it becomes apparent that the frequency response of the analysis-synthesis system is that of a cascade of $(M-1)$ first-order all-pass filters. Consequently, the phase response of the analysis-synthesis system is no longer linear and it becomes necessary to correct the phase using an additional filter operation. In~\cite{gulzow}, for example, a non-recursive filter having an impulse response that is a time-limited, time-inverted impulse response of the analysis-synthesis filter bank is used for correcting the phase. Alternatively, lower-order recursive group-delay equalizers~\cite{antoniou} that approximate the inverse group-delay of the cascade of $(M-1)$ all-pass filters may also be utilized. 

\section{Simulation Results} 
In this section, we show the effectiveness of the proposed method by comparing it with two variants of existing methods, Method A and Method B. We compare their performance for three different types of reference signals: white noise, colored noise and speech.

For Method A, we design the prototype analysis filter using the method described in~\cite{nordholm}. In this method, the filter is designed by simultaneously minimizing the mean-square error in the passband together with the inband aliasing power in the the subband with the widest bandwidth. The desired passband response is constrained to be linear phase with a magnitude of unity. 

The prototype synthesis filter is designed using a modified optimization algorithm where the cost function in (\ref{eq:21}) is replaced with the one in~\cite{nordholm}, given by 
\begin{equation}
\begin{split}
\hat{\Upsilon} & = \sum_{n=0}^{N-1} \sum_{i=0}^{M-1} \sum_{d=1}^{D_i-1}   \left|  G_i(e^{j\omega_n})  H_i(e^{j\omega_n}W_D^d) \right|^2
\end{split}
\label{modCostFunction}
\end{equation}
However, unlike the synthesis design algorithm in~\cite{nordholm}, we do not impose any linear phase constraint in the synthesis filter design for our Method A, since it reduces the degrees of freedom during optimization thereby reducing the performance of the filter even further. At the same time, we also extend the cost function to incorporate variable decimation factors across the subbands. 

For Method B, we design the analysis prototype filter by maximizing the SAR only for the subband with the largest bandwidth. For $\mu > 0$, the general optimization equation for obtaining the analysis filter design in Method B is given by
\begin{eqnarray}
\mbox{minimize } & &  (\sigma_{M/2}^{(a)})^2  \label{genOptmMetB} \\
\mbox{subject to: } & &  \sigma_{M/2}^2 = \mbox{constant} \notag 
\end{eqnarray} 
Method B essentially demonstrates the performance that can be attained when only the largest subband is considered, as was done in~\cite{nordholm}, or when uniform filter bank design methods are employed. For the synthesis prototype filter design, we use the same optimization algorithm as in Section V.

We compare the proposed method with Method A and Method B for two filter-bank design specifications: \newline
(a) Specification 1: $M=16$, $\mu = 0.5$, and $D=2$ and \newline
(b) Specification 2: $M=16$, $\mu = 0.5$, and $D_i$ $\in$ $\{$8, 8, 8, 4, 4, 4, 2, 2, 2, 2, 2, 4, 4, 4, 8, 8$\}$. \newline
We select $\mu = 0.5$ as it closely approximates the Bark frequency scale~\cite{gulzow}. Furthermore, for the ERLE performance comparison experiments in this paper, the adaptive-filter weights are initially set to zero and the adaptation process is started 1 second after the application of the reference and desired signal. This is done so that the error-signal power obtained during the first 1 second can be normalized to 0 dB in the ERLE plots. 

Parameters  $\Omega_l^{(i)}$ and $\Omega_h^{(i)}$, required for computing the analysis filter cost function in (\ref{anaCostFun}), are obtained after solving the line search equation in (\ref{lineSearch}). Their computed values for Specification 1 and Specification 2 are listed in Table \ref{spec0}. It should be noted that the values listed in the table are not unique but have a period of $2\pi$.
 \begin{table}[htb]
\begin{center}
\caption{Values of $\Omega_l^{(i)}$ and $\Omega_h^{(i)}$ for Specification 1 and Specification 2 }
\label{spec0}
{\footnotesize{
\begin{tabular}{||c|c|c|c|c||} \hline \hline
Frequency & \multicolumn{2}{||c|}{Specification 1} & \multicolumn{2}{|c||}{Specification 2} \\ 
\cline{2-5}
bin & $\Omega_l^{(i)}$  & $\Omega_h^{(i)}$  & $\Omega_l^{(i)}$  & $\Omega_h^{(i)}$ \\
\hline \hline
1 & -3.1416  &  3.1416  & -3.1416  &  3.1416 \\
2 & -4.3500  &  1.9331  & -4.5087  &  1.7745 \\ 
3 & -5.2023  &  1.0808  & -5.8933  &  0.3900 \\ 
4 & -5.7960  &  0.4872  & -6.1259  &  0.1574 \\
5 & -6.2832  &  0.0000  & -7.0197  & -0.7365 \\
6 & -6.7703  & -0.4872  & -8.0746  & -1.7914 \\
7 & -7.3639  & -1.0809  & -7.3639  & -1.0809 \\
8 & -8.2163  & -1.9332  & -8.2163  & -1.9332 \\
9 & -9.4248  & -3.1416  & -9.4248  & -3.1416 \\
10 & -10.6332 &  -4.3501 & -10.6332  & -4.3501 \\
11 & -11.4855 &  -5.2025 & -11.4855  & -5.2025 \\ 
12 & -12.0792 &  -5.7960 & -23.3414  & -17.0582 \\
13 & -12.5664 &  -6.2832 & -24.3962  & -18.1131 \\
14 & -13.0536 &  -6.7704 & -25.2901  & -19.0069 \\
15 & -13.6472 &  -7.3641 & -50.6555  & -44.3722 \\
16 & -14.4995 &  -8.2163 & -52.0400  & -45.7567 \\
\hline \hline
\end{tabular}
}}
\end{center}
\end{table}

\subsection{Using white noise as reference signal}
In this sub-section, we compare the performance when the reference signal is white noise; therefore, we set $P_{xx}(\omega)$ to unity when designing the filters using the proposed method. As such, when $P_{xx}(\omega) = 1$ we shall refer to the design method as `Proposed-white'. We also assume no knowledge of the average spectrum of the unknown system, and therefore set $|S(\omega)|^2$ to unity for all of the experiments in this paper.

The desired signal, $d(n)$, is white noise convolved with an impulse response of length 200 that is randomly generated from a normal distribution of unit variance. The length of the adaptive filter in each subband varies with the decimation factor and is set to $256/D_i$ for subband $i$. The NLMS algorithm is employed for adapting the adaptive-filter coefficients in each subband. 

\begin{figure}[htb]
\begin{minipage}[b]{1.0\linewidth}
  \centering
  \centerline{\includegraphics[width=0.9\textwidth]{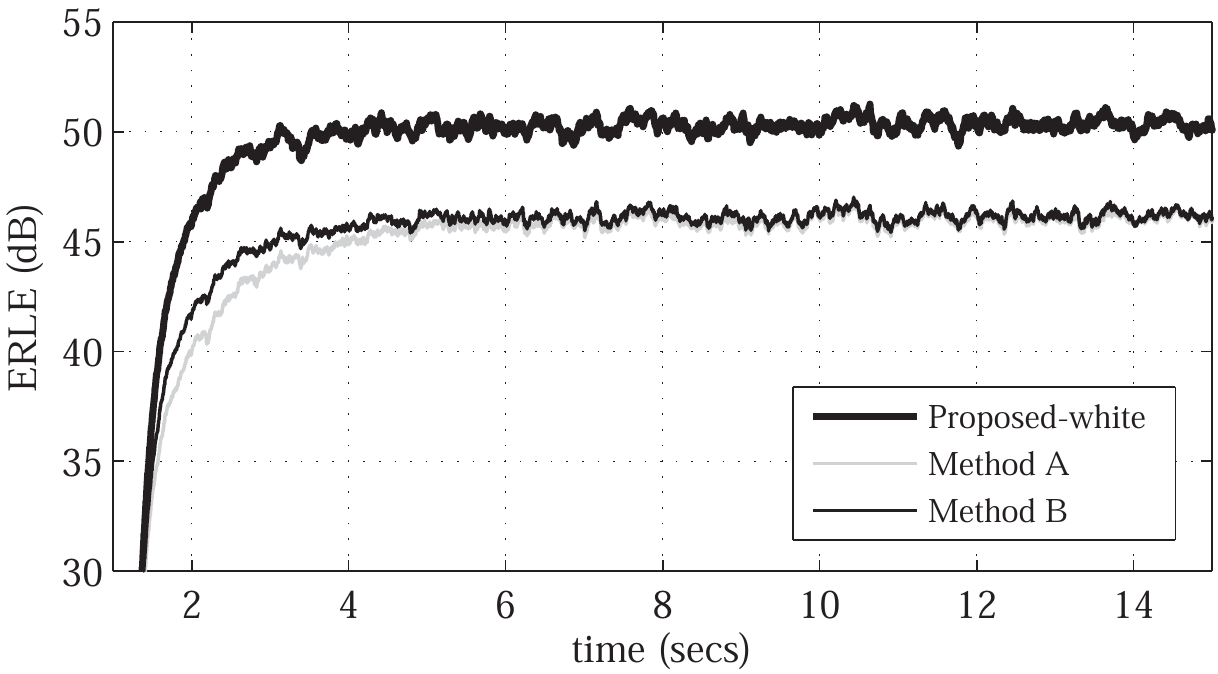}}
  \centerline{(a)}\medskip
\end{minipage}
\begin{minipage}[b]{1.0\linewidth}
  \centering
  \centerline{\includegraphics[width=0.9\textwidth]{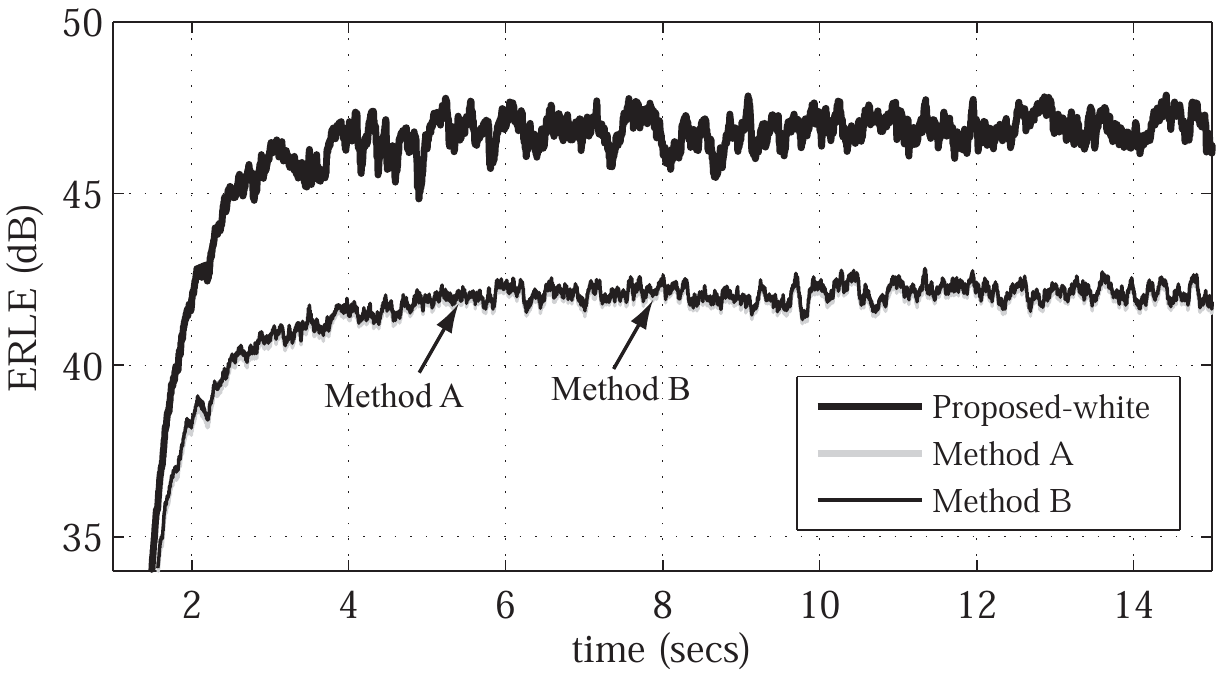}}
  \centerline{(b)}\medskip
\end{minipage}
\caption{Comparison plots of the ERLE as the adaptive filters converges when the reference signal is white noise for (a) Specification 1: $\mu = 0.5$, $M=16$, and $D=2$ (b) Specification 2: $\mu = 0.5$, $M=16$, and $D_i$ $\in$ $\{8$, 8, 8, 4, 4, 4, 2, 2, 2, 2, 2, 4, 4, 4, 8, 8$\}$.}
\label{erlewhite1}
\end{figure}

\begin{figure}[htb]
\begin{center}
\includegraphics[width=0.45\textwidth]{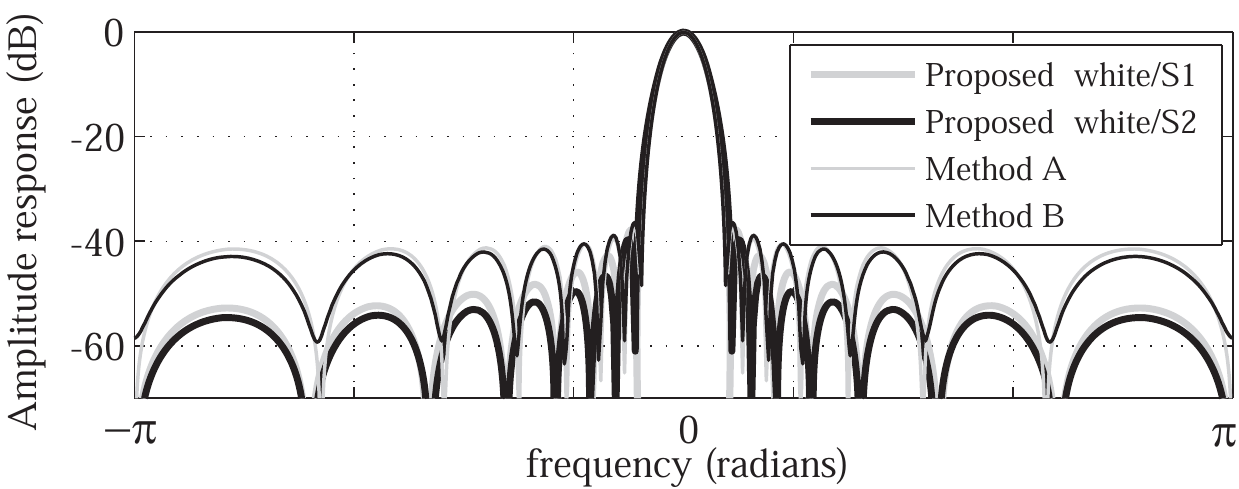}
\caption{Plots of amplitude responses of the analysis prototype filters for the first sub band. S1 and S2 in the figure refer to Specification 1 and Specification 2, respectively.}
\label{anaAmpResp}
\end{center}
\end{figure}

\begin{figure}
\begin{minipage}[b]{1.0\linewidth}
  \centering
  \centerline{\includegraphics[width=0.9\textwidth]{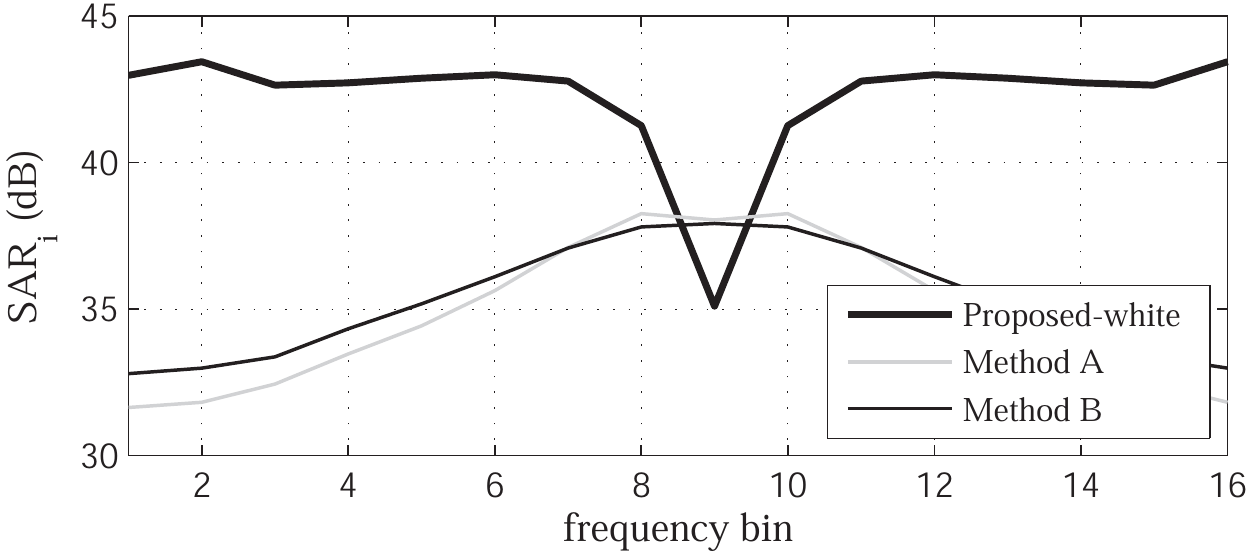}}
  \centerline{(a)}\medskip
\end{minipage}
\begin{minipage}[b]{1.0\linewidth}
  \centering
  \centerline{\includegraphics[width=0.9\textwidth]{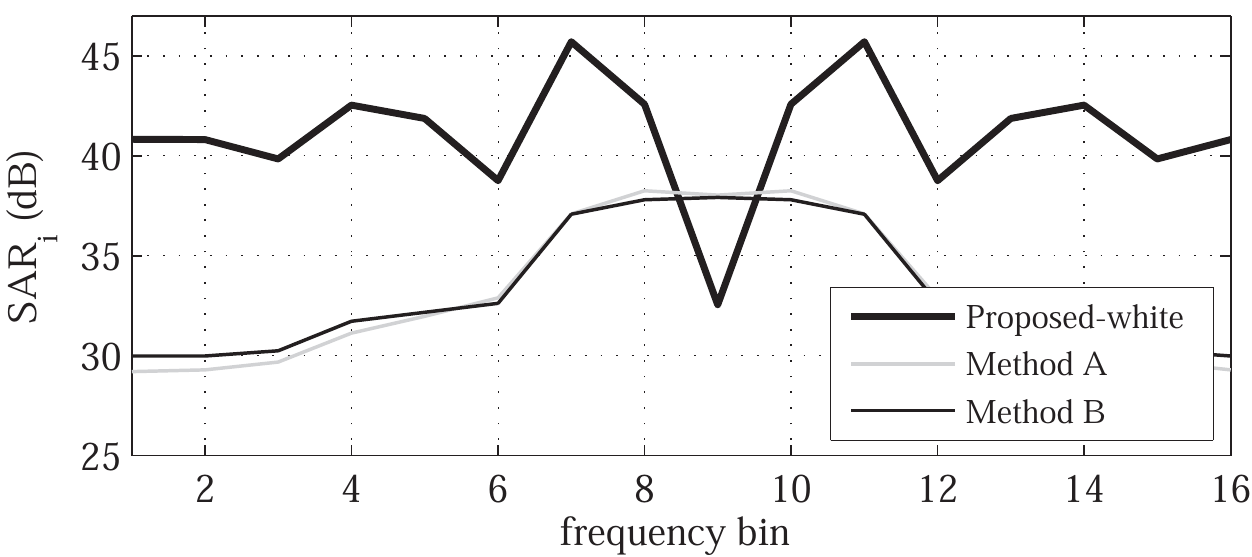}}
  \centerline{(b)}\medskip
\end{minipage}
\caption{Comparison plots of the sub-band SAR for (a) Specification 1 (b) Specification 2.}
\label{sar1}
\end{figure}

The ERLE plot for the two filter bank designs are shown in Figs.~\ref{erlewhite1}(a) and (b) with the corresponding steady-state values tabulated in Table II. As can be seen, the proposed method results in an improvement of several dBs over Method A and Method B. 
\begin{table}[htb]
\begin{center}
\caption{Comparison of the steady-state ERLE}
\label{tab1:1:1}
{\footnotesize{
\begin{tabular}{||c|c|c|c||} \hline \hline
Design & Proposed & Method A & Method B \\
Cases &  (dB) & (dB)  & (dB)  \\ \hline
\hline
 &  &  & \\ 
Spec 1  & 50.34 & 45.99 & 46.16 \\ 
 &  &  & \\ \hline
 &  &  & \\ 
Spec 2  & 46.91 & 41.99 & 42.01 \\  
&  &  & \\  \hline
\end{tabular}
}}
\end{center}
\end{table}
Next, we show comparative plots for the amplitude responses of the prototype analysis filters in Fig.~\ref{anaAmpResp}. 
Then, the full-band SARs computed using (\ref{overallSAR}) are tabulated for the three methods in Table III. Comparing the values in Table II and Table III we observe that the SAR values are about 10 dB smaller than the corresponding ERLE values, but vary proportionally to the ERLE values. The difference between the ERLE and SAR values arises because the SAR in (\ref{overallSAR}) is computed right after analysis whereas the ERLE is estimated after analysis and synthesis. The additional aliasing signal suppression by the synthesis filters results in higher ERLE values that are proportional to the respective SAR values. We then use (\ref{subbandSAR}) to compute the corresponding sub-band SARs, $\mbox{SAR}_i$, which are plotted in Figs.~\ref{sar1}(a) and (b). From the plots, it is apparent that the filters designed using the proposed method have higher sub-band SAR in all the other sub-bands, except in bin 9, which corresponds to the highest frequency sub-band.  The improvement in sub-band SAR in the other sub bands at the expense of a decrease in the highest sub band is not undesirable in acoustic echo cancellation where cancellation of the lower frequency echoes is usually most critical.

\begin{table}[htb]
\begin{center}
\caption{Comparison of the overall SAR}
\label{tab1:1:1}
{\footnotesize{
\begin{tabular}{||c|c|c|c||} \hline \hline
Design & Proposed & Method A & Method B \\
Cases &  (dB) & (dB)  & (dB)  \\ \hline
\hline
 &  &  & \\ 
Spec 1  & 39.00 & 35.84 & 36.23 \\ 
 &  &  & \\ \hline
 &  &  & \\ 
Spec 2 & 38.89 & 32.35 & 32.72 \\  
 &  &  & \\  \hline
\end{tabular}
}}
\end{center}
\end{table}

\begin{figure}[htb]
\begin{minipage}[b]{1.0\linewidth}
  \centering
  \centerline{\includegraphics[width=0.9\textwidth]{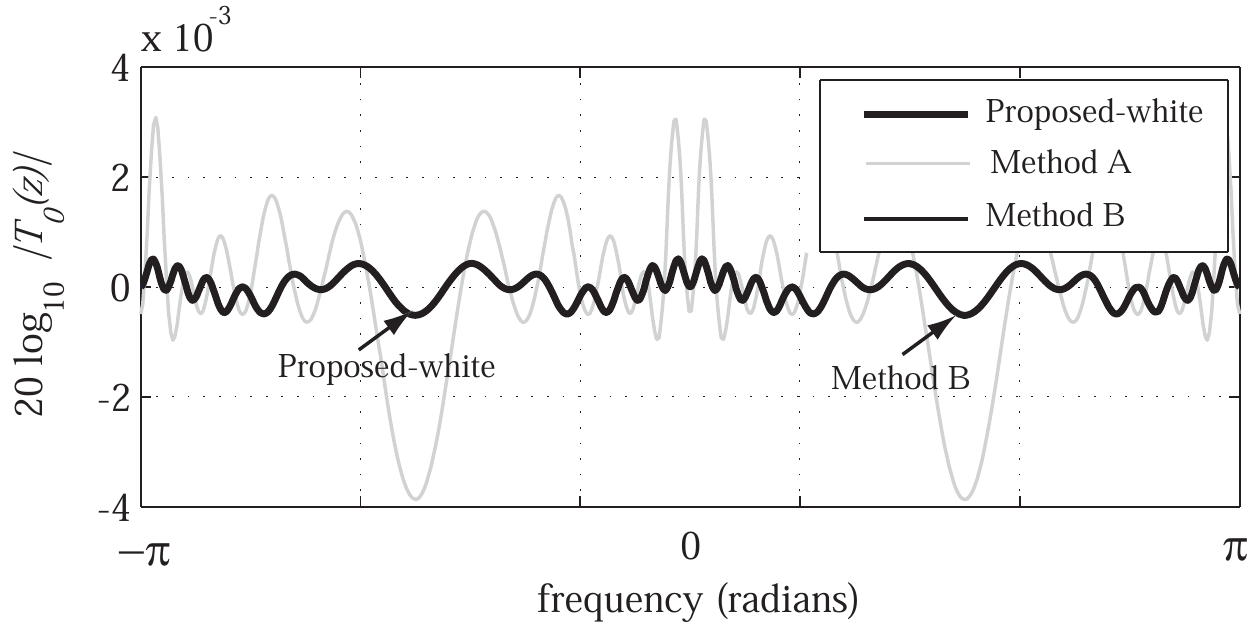}}
  \centerline{(a)}\medskip
\end{minipage}
\begin{minipage}[b]{1.0\linewidth}
  \centering
  \centerline{\includegraphics[width=0.9\textwidth]{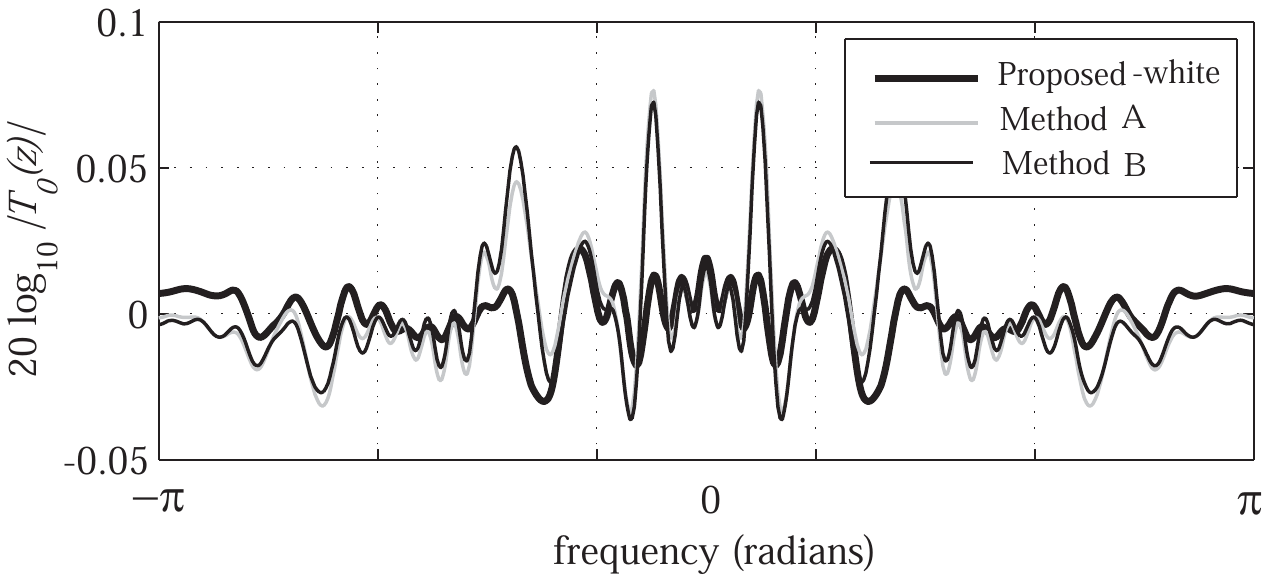}}
  \centerline{(b)}\medskip
\end{minipage}
\caption{Comparison plots of the overall response of the analysis-synthesis system for (a) Specification 1 (b) Specification 2.}
\label{anaSynAmpResp1}
\end{figure}

In Figs.~\ref{anaSynAmpResp1}(a) and (b), we compare the overall amplitude response of the analysis-synthesis system. As can be seen in Fig.~\ref{anaSynAmpResp1}(a), for Specification 1 the proposed method and Method B have the smallest deviation and are also identical. The reason for the identical response is given in Appendix of the paper. For Specification 2, however, the proposed method has the least deviation, even better than Method B. Since the synthesis filter design algorithm for the proposed method and method B are identical, we can conclude that the better overall response in the proposed method is due to better analysis prototype filters. 

\begin{figure}[htb]
\begin{minipage}[b]{1.0\linewidth}
  \centering
  \centerline{\includegraphics[width=0.9\textwidth]{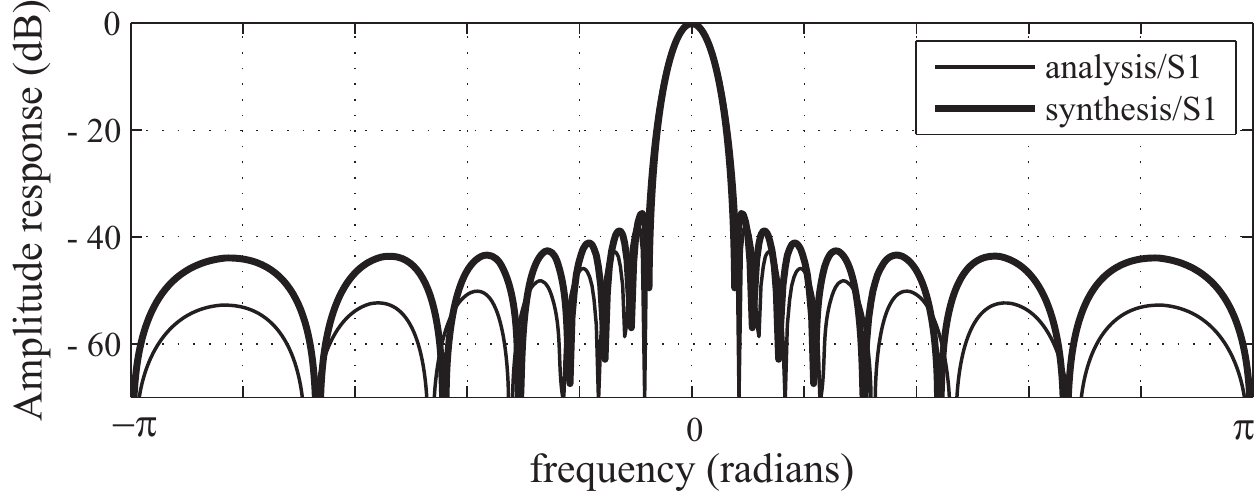}}
  \centerline{(a)}\medskip
\end{minipage}
\begin{minipage}[b]{1.0\linewidth}
  \centering
  \centerline{\includegraphics[width=0.9\textwidth]{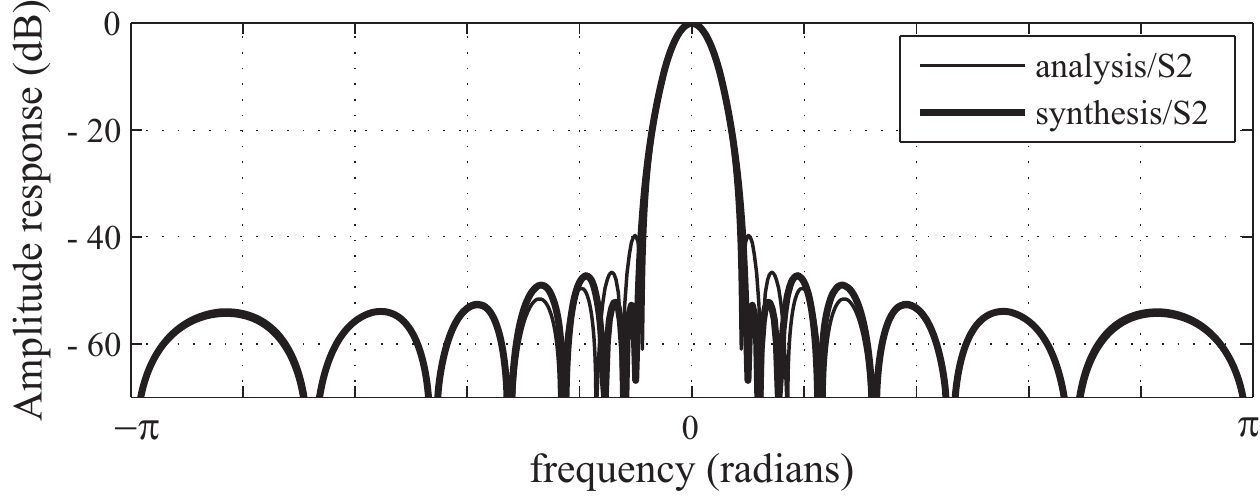}}
  \centerline{(b)}\medskip
\end{minipage}
\caption{Amplitude responses of the proposed analysis and synthesis filters for the first sub-band for (a) Specification 1 (b) Specification 2.}
\label{propAnaSynAmpResp}
\end{figure}

For the proposed method, the filter coefficients of the analysis and synthesis prototype filters for Specification 1 and Specification 2 are given in Table \ref{spec1} and \ref{spec2}, respectively, and their corresponding amplitude responses are shown in Figs.\ref{propAnaSynAmpResp}(a) and (b).
 \begin{table}[htb]
\begin{center}
\caption{Filter coefficients of the analysis and synthesis prototype filters for Specification 1}
\label{spec1}
{\footnotesize{
\begin{tabular}{||c|c||} \hline \hline
Analysis filter & Synthesis filter \\
\hline \hline
0.006784387784996 & 0.010741021222032 \\
0.017946339940064 & 0.021926742080588 \\
0.033423742747281 & 0.037203345945288 \\
0.052593030750424 & 0.054888469644485 \\
0.073281609489174 & 0.073198870790831 \\
0.092560997189968 & 0.089930206035430 \\
0.107586665017076 & 0.102631178713128 \\
0.115769055764595 & 0.109536252754658 \\
0.115773770718600 & 0.109531791827139 \\
0.107604509223835 & 0.102614159240595 \\
0.092582338907088 & 0.089909475677339 \\
0.073306631225949 & 0.073173885835641 \\
0.052609833237766 & 0.054870939407773 \\
0.033437308233337 & 0.037188252580875 \\
0.017951793881796 & 0.021920080508076 \\
0.006787985888052 & 0.010735327736125 \\
\hline \hline
\end{tabular}
}}
\end{center}
\end{table}
 \begin{table}[htb]
\begin{center}
\caption{Filter coefficients of the analysis and synthesis prototype filters for Specification 2}
\label{spec2}
{\footnotesize{
\begin{tabular}{||c|c||} \hline \hline
Analysis filter & Synthesis filter \\
\hline \hline
0.005575845754339 & 0.005485648407843 \\
0.015692207827265 & 0.014752288378833 \\
0.031111735472651 & 0.028969305498630 \\
0.050989625731218 & 0.048587305437593 \\
0.073001594213692 & 0.071841335493898 \\
0.093920111506342 & 0.094628233023347 \\
0.110318424188072 & 0.112800041474789 \\
0.119330344129408 & 0.122930412131727 \\
0.119335069580386 & 0.122932039416339 \\
0.110340087229245 & 0.112799214437691 \\
0.093941237149871 & 0.094632898386205 \\
0.073026217924965 & 0.071844987310398 \\
0.051009071444013 & 0.048590076561173 \\ 
0.031126188682438 & 0.028968967987558 \\
0.015699544319015 & 0.014751755969681 \\
0.005582694847079 & 0.005485490084296 \\
\hline \hline
\end{tabular}
}}
\end{center}
\end{table}

\subsection{Using colored noise as reference signal}
\begin{figure}[tbp]
\begin{center}
\includegraphics[width=0.45\textwidth]{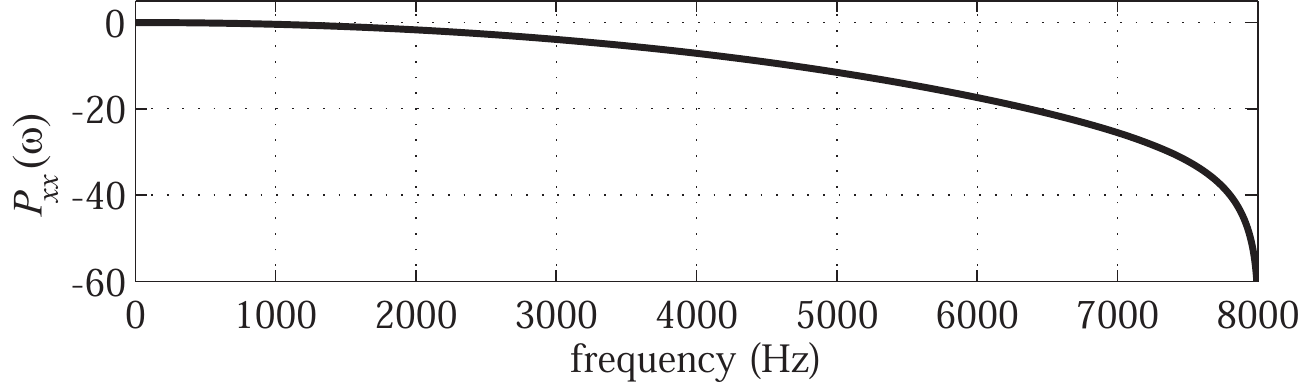}
\caption{Power spectrum of the colored noise.}
\label{colordNoiseSpec}
\end{center}
\end{figure}

In this subsection, we compare the ERLE performance  when the reference signal is colored noise. We therefore design a second set of analysis and synthesis prototype filters that takes the spectrum of the reference signal, $P_{xx}(e^{j\omega})$, into account when designing the filters for Specification 1 and Specification 2. That is, we set $P_{xx}(e^{j\omega})$ in (\ref{model}) as the power spectrum of the colored noise. A plot of the spectrum is shown in Fig.~\ref{colordNoiseSpec}. To differentiate from the design in Subsection VI-A where $P_{xx}(e^{j\omega})$ is unity, we shall refer to this design method as `Proposed-colored'.
The colored noise is obtained by passing the signal through a low-pass FIR filter of order 5; it is estimated to have an eigenvalue spread of 126. Like in Subsection VI-A, the desired signal is obtained by convolving the colored noise with a randomly generated impulse response of length 200. 

\begin{figure}[tpb]
\begin{minipage}[b]{1.0\linewidth}
  \centering
  \centerline{\includegraphics[width=0.9\textwidth]{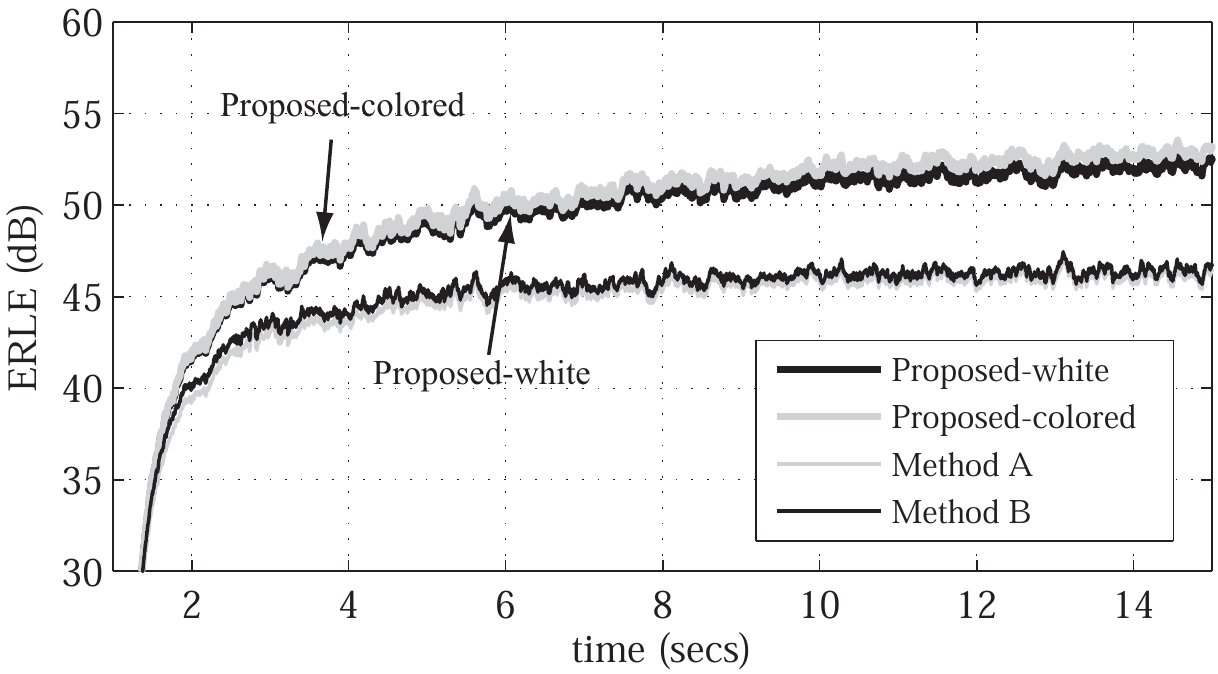}}
  \centerline{(a)}\medskip
\end{minipage}
\begin{minipage}[b]{1.0\linewidth}
  \centering
  \centerline{\includegraphics[width=0.9\textwidth]{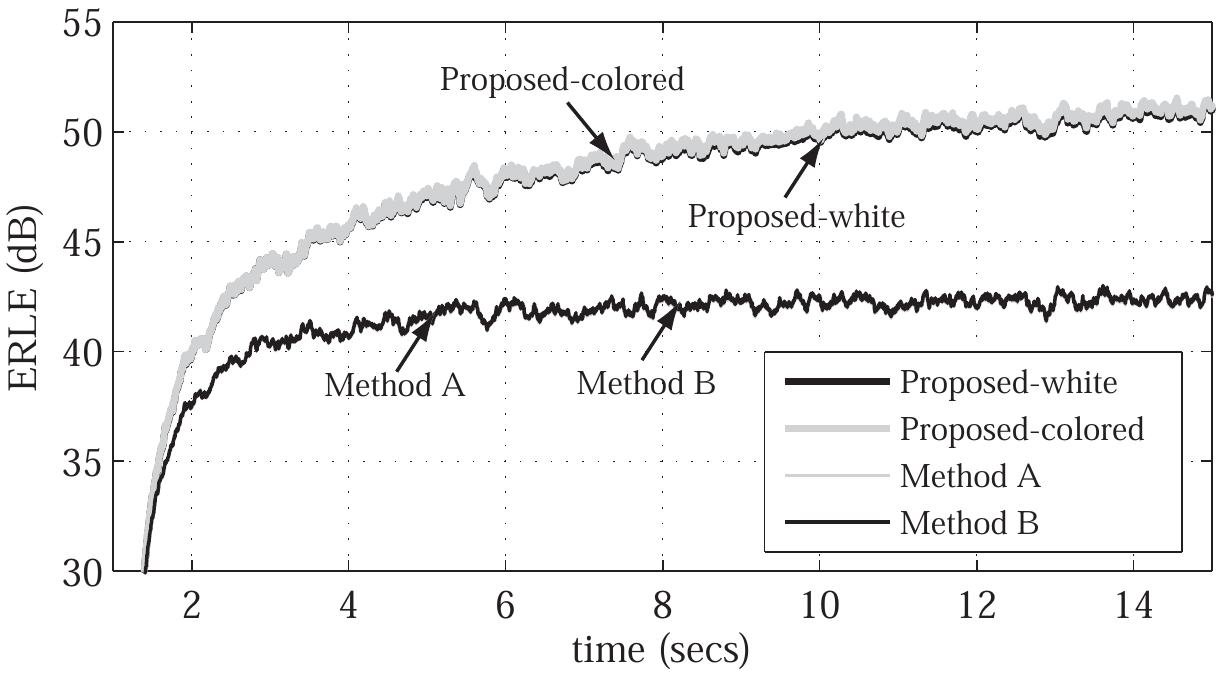}}
  \centerline{(b)}\medskip
\end{minipage}
\caption{Comparison plots of the ERLE as the adaptive filters converges when the reference signal is colored noise for (a) Specification 1 (b) Specification 2}
\label{erlecolored}
\end{figure}

The ERLE plot for the two filter bank designs are shown in Figs.~\ref{erlecolored}(a) and (b). As can be seen, the proposed methods show an improvement of several dBs over Method A and Method B. It is interesting to note that for Specification 2, the improvement of the `Proposed-color' method over the `Proposed-white' method is not as high as in Specification 1. This is because the constraints imposed by the higher decimation factors in Specification 2 limits the degree of freedom in the minimization of the aliasing power for a certain change in $P_{xx}(e^{j\omega})$.

\subsection{Using speech as reference signal}
In this subsection, we compare the ERLE performance when the reference signal is speech. As in Subsection VI-B, we design a second set of analysis and synthesis prototype filters where $P_{xx}(e^{j\omega})$ is set to the average power spectrum of speech. We refer to this design method as `Proposed-speech'.

\begin{figure}[tbp]
\begin{center}
\includegraphics[width=0.45\textwidth]{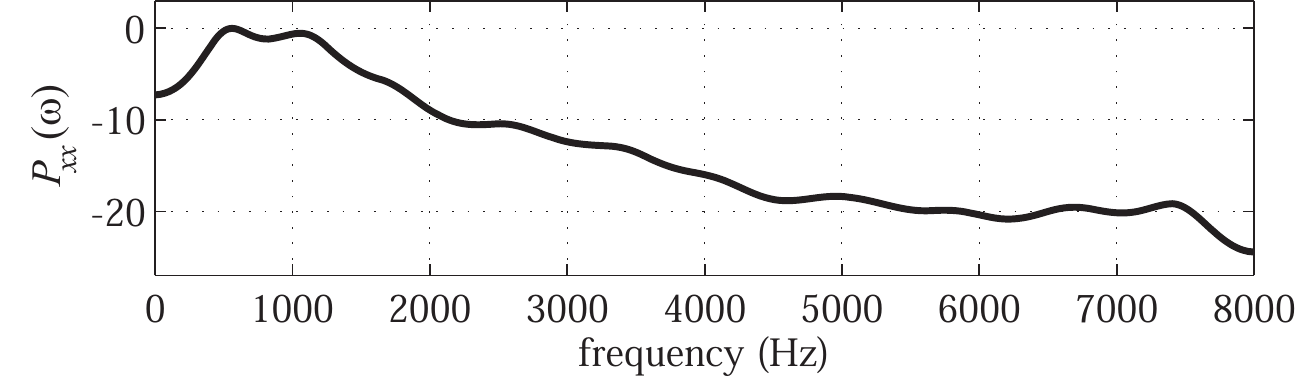}
\caption{Average power spectrum of the speech signal.}
\label{speechSpec}
\end{center}
\end{figure}

\begin{figure}[tbp]
\begin{center}
\includegraphics[width=0.45\textwidth]{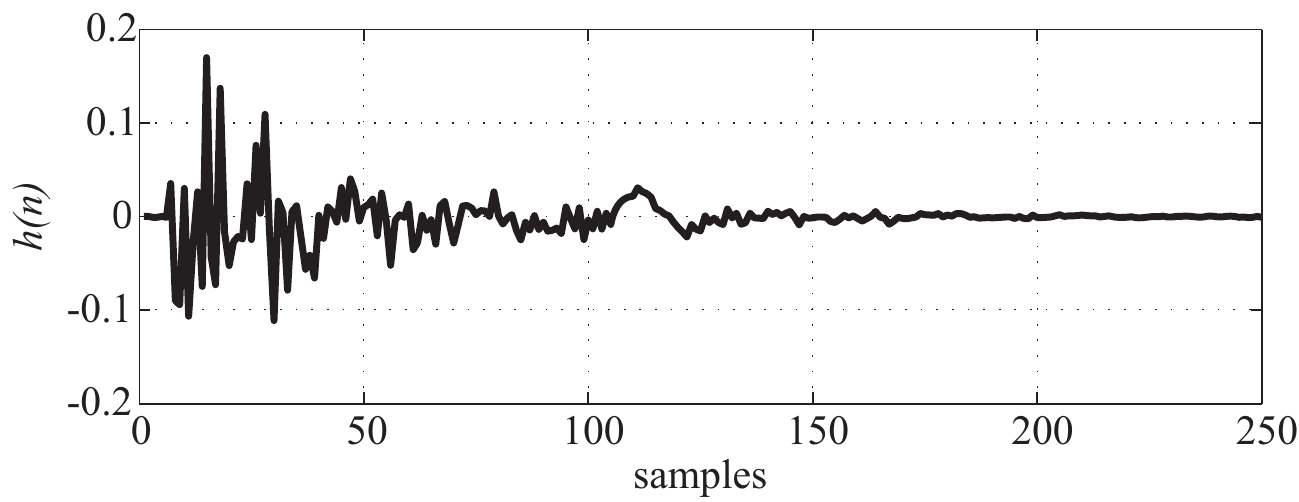}
\caption{Impulse response measured in a compact-size car at Nyquist frequency of 8 kHz.}
\label{carImpResp}
\end{center}
\end{figure}

To compute the average power spectrum of speech, we took speech signals of 3 males and 3 females speakers from the ATIS database~\cite{atis} and computed their average spectrum, which is plotted in Fig.~\ref{speechSpec}. The duration of the signal is about 5 minutes with a Nyquist frequency of 8 kHz. To avoid including the silence portion of speech when computing the average, we use a simple energy detector to make the classification. 
Unlike the experiments in the previous subsection where we used a randomly generated impulse response, in this section we use a real impulse response, measured in a compact-sized car, to generate the desired signal from the reference speech signal; a plot of the impulse response is shown in Fig.\ref{carImpResp}.

\begin{figure}[htb]
\begin{minipage}[b]{1.0\linewidth}
  \centering
  \centerline{\includegraphics[width=0.95\textwidth]{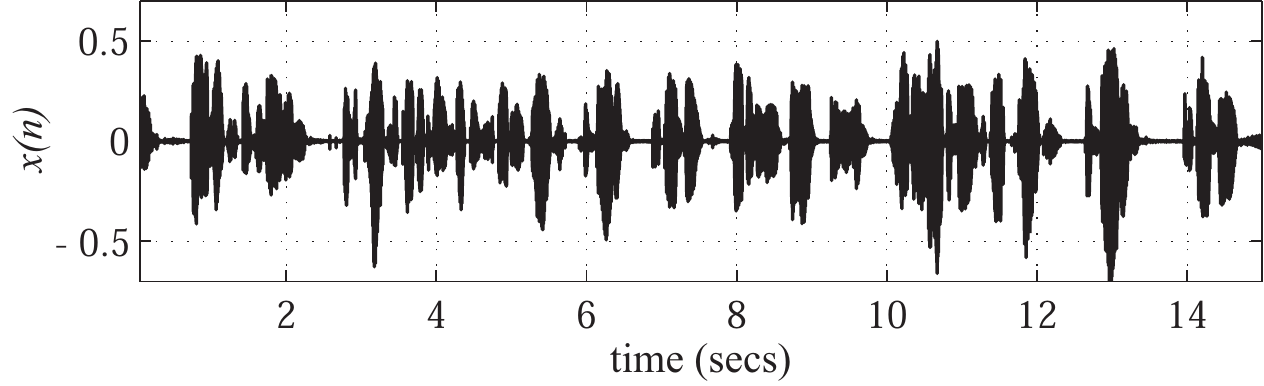}}
  \centerline{(a)}\medskip
\end{minipage}
\begin{minipage}[b]{1.0\linewidth}
  \centering
  \centerline{\includegraphics[width=0.95\textwidth]{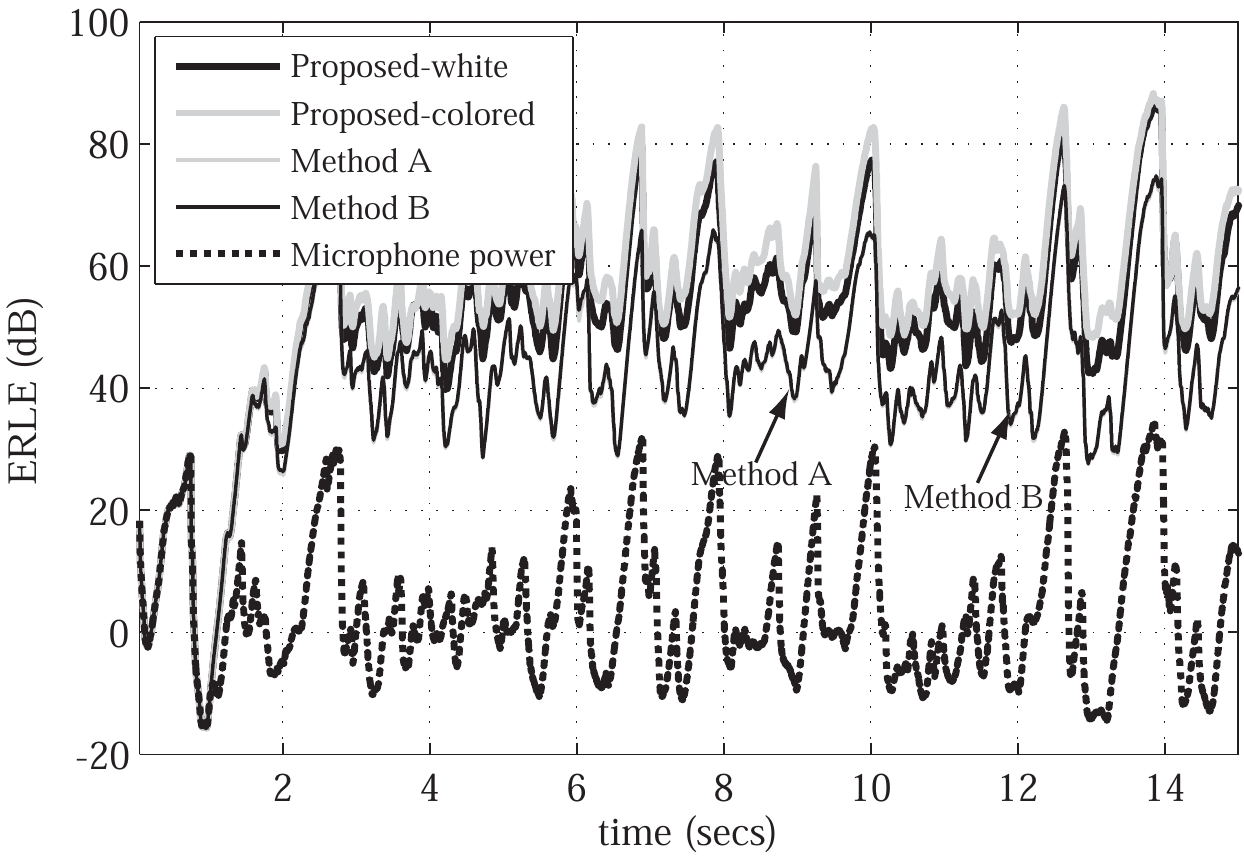}}
  \centerline{(b)}\medskip
\end{minipage}
\begin{minipage}[b]{1.0\linewidth}
  \centering
  \centerline{\includegraphics[width=0.95\textwidth]{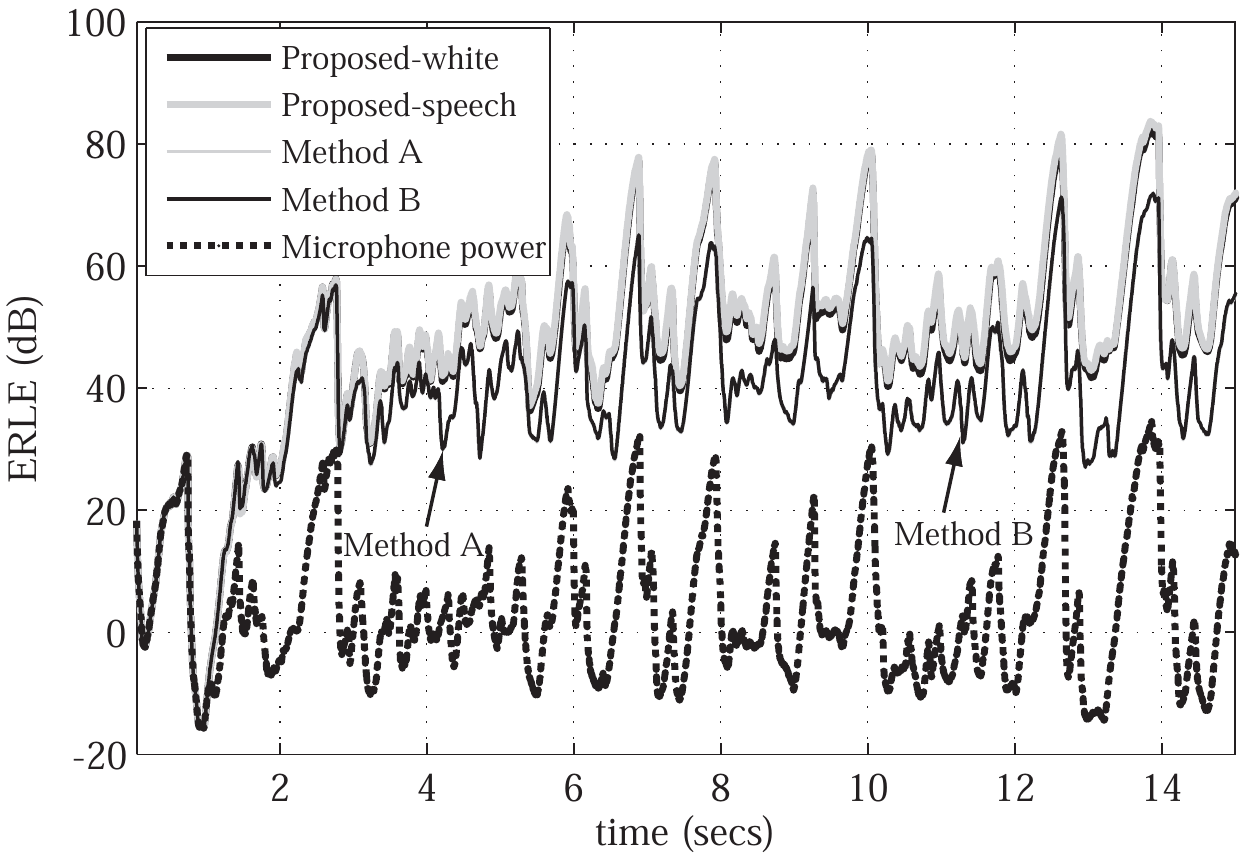}}
  \centerline{(c)}\medskip
\end{minipage}
\caption{Comparison plots of the ERLE for (b) Specification 1 (c) Specification 2, when the reference signal is speech. The upper plot is the reference speech-signal to the adaptive filter. In the plots, the adaptation of the adaptive filter weights is started after 1 seconds.}
\label{erlespeech}
\end{figure}

The reference speech signal to the adaptive filter is shown in Fig.\ref{erlespeech}(a) and the ERLE plot for the two filter bank designs are shown in Figs.~\ref{erlespeech}(b) and (c). As can be seen, the proposed methods show improvements of several dBs over Method A and Method B. And, like in Subsection VI-B, the improvement of the `Proposed-speech' method over the 'Proposed-white' method is higher for Specification 1.

It should be noted that for the sake of comparison, we have used the NLMS algorithm with a fixed step size in our experiments. However, in practical applications the convergence rate of the adaptive filters can be significantly improved by employing various techniques~\cite{gerhard}, such as varying the step-sizes as the adaptation progresses, or using more powerful adaptation algorithms like the improved-PNLMS~\cite{benesty} or the affine projection algorithms~\cite{makino}.  
 
\section{Conclusions}
A new method for designing non-uniform filter-banks for acoustic echo cancellation has been described. In the method, the analysis prototype filter is framed as a convex optimization problem that maximizes the SAR in the analysis banks. Since each subband has different bandwidth, the contribution to the overall SAR from each subband is taken into account during optimization. To increase the degrees of freedom during optimization no constraints are imposed on the phase of the filter. And to ensure low delay, the filter is constrained to be minimum phase. Experimental results show that the proposed method results in filter banks with fast convergence and superior ERLE when compared to filter banks designed using existing methods. 
\section*{Acknowledgment}
The authors are grateful to the Natural Sciences and Engineering Research Council of Canada for supporting this work.
\section*{Appendix}
In this appendix, we show that if the decimation factors across the sub-bands are the same and the analysis filter used in deriving the synthesis filter in (\ref{eq:22}) has no zero coefficients, the analysis-synthesis amplitude response is, up to a scale factor, independent of the analysis prototype filter.

Setting the decimation factors to be equal across the subbands in (\ref{eq:20}) we get
\begin{equation}
\begin{split}
T_a(e^{j\omega}, l) & = \sum_{i=0}^{M-1} G_i(e^{j\omega}) \sum_{d=1}^{D-1} W_{D}^{-dl}   H_i(e^{j\omega} W_{D}^d)
\end{split}
\label{AliasingSynSameD}
\end{equation}
Upon expanding $H_i(e^{j\omega} W_{D}^d)$ and $G_i(e^{j\omega})$, interchanging the summations and simplifying we obtain
\begin{equation}
\begin{split}
T_a(e^{j\omega}, l) &  = \sum_{d=1}^{D-1} W_{D}^{-dl}  \sum_{n=0}^{M-1} m(n) A(e^{j\omega})^{M-n-1} A(e^{j\omega} W^d_D)^n
\end{split}
\label{AliasingSynSameDSimple}
\end{equation}
where
\begin{equation}
m(n) = g(n)h(n),
\label{dotProduct}
\end{equation}
$g(n)$ is the optimization variable and $h(n)$ is known. Therefore, from (\ref{dotProduct}) it is apparent that if $h(n)$ is not zero, $m(n)$ in (\ref{AliasingSynSameDSimple}) remains unconstrained, and, consequently, the minimization of the the cost function in (\ref{eq:21}) under the constraint that $\mathbf{h}^T\mathbf{g} = \sum_n m(n) = 1$, is independent of the analysis prototype filter. If, however, $h(n)$ is zero for $n=k$, then $m(k)$ is also constrained to zero, and $g(k)$ can have arbitrary values. Because of this scenario, we introduce the regularization term $\delta \mathbf{g}^T \mathbf{g}$ in the optimization problem in (\ref{eq:22}) so that solution of $g(n)$ with the minimum L2 norm is always selected.


\begin{thebibliography}{99}  \setlength{\itemsep}{-1pt}
\footnotesize{
\bibitem{zwicker} E. Zwicker and H. Fastl,
{\em Psychoacoustics - Facts and Models}, Second Edition, Springer 1999.
\bibitem{oppenheim} A. V. Oppenheim, D. Johnson, and K. Steiglitz,
``Comparison of spectra with unequal resolution using Fast Fourier Transform,''
{\em Proc. of the IEEE}, vol. 59, no. 2, pp. 299-301, Feb. 1971.

\bibitem{vary1} P. Vary, ``Digital Filter Banks with Unequal Resolution,'' {\em Short Communication Digest of European Signal Processing Conf. (EUSIPCO)}, Lausanne, Switzerland, Sept. 1980, pp. 41-42.

\bibitem{doblinger} G. Doblinger, ``An Efficient Algorithm for Uniform and Nonuniform
Digital Filter Banks,'' {\em Proc. of Intl. Symp. on Circuits and Systems (ISCAS)}, Singapore, June 1991, vol. 1, pp. 646-649.

\bibitem{gulzow} T. Gulzow, A. Engelsberg, and U. Heute,
``Comparison of a discrete wavelet transformation and a nonuniform polyphase filter-bank applied to spectral-subtraction speech enhancement,''
{\em Signal Processing, Elsevier}, vol. 64, no. 1, pp. 5-19, Jan. 1998.

\bibitem{nordholm} J. M. de Hann, N. Grbic, I. Claesson, and S. Nordholm, ``Design and evaluation of nonuniform DFT filter banks in subband microphone arrays,''
{\em Proceedings of ICASSP 2002}, vol. 2, pp. 1173-1176.

\bibitem{gerhard} E. Hansler and G. Schmidt,
{\em Acoustic echo and noise control - A practical approach}, Wiley-Interscience 2004.

\bibitem{princen} J. Princen, ``The Design of Nonuniform Filter Banks,'' {\em IEEE Transactions on Signal Processing}, vol. 43, no. 11, pp. 2550-2560, November 1995.

\bibitem{dumitrescu} B. Dumitrescu, R. Bregovic, T. Saramaki, ``Design of low-delay nonuniform oversampled filterbanks,'' {\em Signal Processing}, vol.88, pp. 2518-2525, 2008.
 
\bibitem{cvetkovic} Z. Cvetkovic and J. D. Johnston, ``Nonuniform oversampled
filter banks for audio signal processing,'' {\em IEEE Transactions On Speech and Audio Processing}, vol. 11, no. 5, pp. 393-399, 2003.

\bibitem{mccloud1} M. L. McCloud and D. M. Etter, ``Subband adaptive filtering with
time-varying nonuniform filter banks,'' {\em Proc. IEEE Int. Conf. Acoust., Speech, Signal Process.}, vol. 3, 1997, pp. 1953-1956.

\bibitem{griesbach2} J. D. Griesbach, T. Bose, and D. M. Etter, ``Non-uniform filterbank
bandwidth allocation for system modeling subband adaptive filters,'' {\em Proc. IEEE Int. Conf. Acoust., Speech, Signal Process.}, vol. 3, 1999, pp. 1473-1476.

\bibitem{lee} J-J. Lee and B. G. Lee,
``A Design of non-uniform cosine modulated filter banks,''
{\em IEEE Trans. Circuits Syst. II, Analog Digit. Signal Process.}, vol. 42, no. 11, pp. 732-737, Nov. 1995.

\bibitem{batalheiro} M. R. Petraglia and P. B. Batalheiro, ``Nonuniform subband adaptive filtering with critical sampling,'' {\em IEEE Trans. Signal Process.}, vol. 56, no. 2, pp. 565-575, Feb. 2008.

\bibitem{petraglia2} M. R. Petraglia, R. G. Alves, and P. S. R. Diniz, ``New structures for adaptive filtering in subbands with critical sampling,'' {\em IEEE Trans. Signal Process.}, vol. 48, no. 12, pp. 3316-3327, Dec 2000.

\bibitem{kellerman} W. Kellermann, ``Analysis and design of multirate systems for cancellation of acoustical echoes,'' {\em Proc. IEEE Int. Conf. Acoust., Speech, Signal Process.}, vol. 5, pp. 2570-2573, 1988.

\bibitem{hart} M. Harteneck, S. Weiss, and R. W. Stewart,
``Design of near perfect reconstruction oversampled filter banks for subband adaptive filters,'' {\em IEEE Trans. Circuits Syst. II, Analog Digit. Signal Process.}, vol. 46, no. 8, pp. 1081-1085, Nov. 1999.

\bibitem{vetterli} A. Gilloire and M. Vetterli,``Adaptive filtering in subbands with critical sampling: Analysis, experiments and applications to acoustic echo cancelation,'' {\em IEEE Trans. Signal Processing}, vol. 40, pp. 1862-1875, Aug. 1992.

\bibitem{hansler} E. Hansler, ``The hands-free telephone problem: An annoted bibliography,'' {\em Signal Process.}, vol. 27, no. 3, pp. 259-271, June 1992.

\bibitem{weiss} S. Weiss, R. W. Stewart, A. Stenger, and R. Rabenstein, ``Steady-state performance limitations of subband adaptive filters,''
{\em IEEE Trans. Signal Process.}, vol. 49, pp. 1982-1991, Sep. 2001.

\bibitem{wilbur} M. R. Wilbur, T. N. Davidson, and J. P. Reilly, ``Efficient design of oversampled NPR GDFT filterbanks,''
{\em IEEE Trans. Signal Process.}, vol. 52, pp. 1947-1963, Jul. 2004.

\bibitem{somayazulu} V. S. Somayazulu, S. K. Mitra, and J. J. Shynk, ``Adaptive line enhancement using multirate techniques,'' {\em Proc. IEEE Int. Conf. Acoust., Speech, Signal Process.}, vol. 2, pp. 928-931, May 1989.

\bibitem{slock} D. T. M. Slock, ``Fractionally-spaced subband and multiresolution adaptive filters,'' {\em Proc. IEEE Int. Conf. Acoust., Speech, Signal Process.}, vol. 5, 1991, pp. 3693-3696.

\bibitem{petraglia} M. R. Petraglia and S. K. Mitra, ``Performance analysis of adaptive filter structures based on subband decompositions,'' {\em Proc. IEEE Int. Symp. Circuits Syst.}, vol. I, pp. 60-63, 1993.

\bibitem{ohno} S. Ohno and H. Sakai, ``Spectral analysis of subband adaptive digital filters,'' {\em IEEE Trans. Signal Processing}, vol. 48, pp. 254-257, Jan. 2000.

\bibitem{morgan} D. R. Morgan and J. C. Thi, ``A delayless subband adaptive filter architecture,'' {\em IEEE Trans. Signal Processing}, vol. 43, pp. 1819-1830, Aug. 1995.

\bibitem{kliewer} E. Galijasevic and J. Kliewer,
``Design of allpass-based non-uniform oversampled DFT filter banks,''
{\em Proceedings of ICASSP 2002}, vol. 2, pp. 1181-1184.

\bibitem{vary2} H. W. Lollmann, G. Dartmann, and P. Vary,
``Least-squares design of subsampled allpass transformed DFT filter-banks with LTI property,''
{\em Proceedings of ICASSP 2008}, pp. 3529-3532, 2008.

\bibitem{vary3} H. W. Lollmann and P. Vary, ``Least-Squares Design of DFT Filter-Banks Based on Allpass Transformation of Higher Order,'' {\em IEEE Trans. Signal Processing}, vol. 58, no. 4, pp. 2393-2398, Apr. 2010.

\bibitem{vo} B. Vo and S. Nordholm, ``Non-uniform DFT filter bank design with semi-definite programming,'' {\em in Proc. Int. Symp. Signal Processing Information Technology (ISSPIT)}, Darmstadt, Germany, Dec. 2003, pp.42-45.

\bibitem{rnongpiur1} R. C. Nongpiur and D. J. Shpak, ``Bi-Criterion Optimization of Non-Uniform Filter Banks for Acoustic Echo Cancellation,'' {\em 2011 IEEE International Symposium on Circuit and Systems (ISCAS 2011)}, Rio de Janeiro, Brazil.

\bibitem{rnongpiur2} R. C. Nongpiur and D. J. Shpak, ``Maximizing the Signal/Alias Ratio in Non-Uniform Filter Banks for Acoustic Echo Cancellation,'' {\em 9th IEEE International NEWCAS Conference (NEWCAS 2011)}, Bordeaux, France.

\bibitem{algobook} {\em Programs for Digital Signal Processing}, IEEE Press, 1979.

\bibitem{wslu} A. Antoniou, W.-S. Lu, {\em Practical Optimization - Algorithms and engineering applications}, Springer 2007.

\bibitem{antoniou} A. Antoniou, {\em Digital signal processing: signals, systems, and filters}, McGraw-Hill, New York, 2005.

\bibitem{atis} C. Hemphill, J. Godfrey, and G. Doddington,
``The ATIS spoken language system pilot corpus,''
{\em  Proceedings of the DARPA Speech and Natural Language Workshop}, 1984.

\bibitem{benesty} J. Benesty and S. L. Gay, ``An improved PNLMS algorithm,'' {\em Proc. IEEE Int. Conf. Acoust., Speech, Signal Process.}, vol. 2, 2002, pp. 1881-1884.

\bibitem{makino} S. Makino, J. Noebauer, Y. Haneda, and A. Nakagawa, ``SSB subband echo canceller using low-order projection algorithm,'' {\em Proc. IEEE Int. Conf. Acoust., Speech, Signal Process.}, vol. 2, 1996, pp. 945-948.
}
\end{thebibliography}
\end{document}